\documentclass[fleqn,usenatbib]{mnras}

\usepackage{newtxtext,newtxmath}
\usepackage[T1]{fontenc}

\DeclareRobustCommand{\VAN}[3]{#2}
\let\VANthebibliography\thebibliography
\def\thebibliography{\DeclareRobustCommand{\VAN}[3]{##3}\VANthebibliography}

\usepackage{graphicx}	
\usepackage{amsmath}	
\usepackage{xspace}

\newcommand{\fbound}{ \ensuremath{f_\text{bound}}\xspace }	

\title[Hierarchically formed embedded star clusters]{Are hierarchically formed embedded star clusters surviving gas expulsion depending on their initial conditions?}

\author[R. Dom\'inguez et al.]{R. Dom\'inguez$^1$\thanks{E-mail:raul.dominguez@stud.uni-heidelberg.de},
 J.P. Farias$^2$, M. Fellhauer$^3$, Ralf S. Klessen$^{1,4}$.
\\$^1$ Universit\"at Heidelberg, Zentrum f\"ur Astronomie, Institut f\"ur Theoretische Astrophysik, Albert-Ueberle-Stra$\beta$e 2, 69120 Heidelberg, Germany
\\$^2$ Dept. of Space, Earth \& Environment, Chalmers University of Technology, Gothenburg, Sweden
\\$^3$ Departamento de Astronom\'ia, Universidad de Concepci\'on, Casilla 160-C, Concepci\'on, Chile
\\$^4$ Universit\"at Heidelberg, Interdisziplin\"ares Zentrum fur Wissenschaftliches Rechnen, Im Neuenheimer Feld 205, 69120 Heidelberg, Germany
  }
\begin{document}
\date{Accepted -----. Received -----; in original form -----}

\pagerange{\pageref{firstpage}--\pageref{lastpage}} \pubyear{2020}

\maketitle

\label{firstpage}
  
\begin{abstract}
We investigate the dissolution process of young embedded star clusters with different primordial mass segregation levels using fractal distributions by means of $N$-body simulations.
We combine several star clusters in virial and subvirial global states with Plummer and uniform density profiles to mimic the gas. The star clusters have masses of $M_\text{stars}$ = 500 $M_\odot$ which follow an initial mass function where the stars have maximum distance from the centre of $r = 1.5$ pc. The clusters are placed in clouds which at the same radius have masses of $M_\text{cloud}$ = 2000 M$_\odot$, resulting in star formation efficiency of 0.2. We remove the background potential instantaneously at a very early phase, mimicking the most destructive scenario of gas expulsion. The evolution of the fraction of bound stellar mass is followed for a total of 16 Myr for simulations with stellar evolution and without. We compare our results with previous works using equal-mass particles where an analytical physical model was used to estimate the bound mass fraction after gas expulsion.
We find that independent of the initial condition, the fraction of bound stellar mass can be well predicted just right after the gas expulsion, but tends to be lower at later stages, as these systems evolve due to the stronger two-body interactions resulting from the inclusion of a realistic initial mass function. 
This discrepancy is independent of the primordial mass segregation level.
\end{abstract}


\section{Introduction}
\label{sec:intro}
Young star clusters are usually found embedded in molecular clouds from which they were recently born. Their early evolution is dominated by feedback 
processes such as ultraviolet radiation and massive stellar winds from OB stars, or supernovae (SNe) explosions, that eventually remove the natal gas. As gas is expelled, star clusters lose gravitational potential resulting in their dissolution into the field \citep[see e.g..][]{1978A&A....70...57T, 1980ApJ...235..986H,1984BAAS...16..409M, 1997MNRAS.284..785G, 2000ApJ...542..964A, 2001MNRAS.323..988G, 2003MNRAS.338..665B, 2003MNRAS.338..673B, 2005ApJ...630..879F, 2006MNRAS.369L...9B, 2007MNRAS.380.1589B,  2011MNRAS.414.3036S, 2016MNRAS.460.2997L, 2017A&A...600A..49B, 2017ApJ...838..116F, Farias2018, 2017A&A...605A.119S, 2018ApJ...863..171S, 2020IAUS..351..507S}. In most scenarios, the feedback is assumed to be strong enough to disrupt the molecular cloud completely, preventing any further star formation \citep{2011ApJ...729..133M,2010ApJ...709...27W}.

Seminal studies like that of \citet{2007MNRAS.380.1589B} have shown that if gas expulsion happens instantly, e.g., by a SNe explosion, only star clusters with global star formation efficiency (SFE) higher than 0.2 can retain a fraction of bound stars after the gas is gone. 
This SFE limit can decrease if gas expulsion happens over larger timescales since stars have time to adjust to the change in the gravitational potential. However, later studies have indicated that other aspects can also bring down this limit, for instance, if the gas is distributed in a less concentrated form, its contribution to the binding potential of the cluster becomes less important \citep{2017A&A...605A.119S}. Additionally, a wide range of post-gas-expulsion bound fractions can be found if star clusters are formed hierarchically, since the relaxation processes that erase substructure can also raise the effective star formation efficiency within the half mass radius of star clusters by the time gas expulsion happens \citep{2011MNRAS.414.3036S}. 
There are other aspects that have also been ignored so far. For instance, most of the previous studies are based on equal mass particles to exclude the effects of strong dynamical interactions between different mass stars, to isolate the effect of the different processes under study. However, stars have a wide range of masses, and an important aspect that is currently under active debate is whether massive stars are born in preferential locations within the star forming regions \citep{1982NYASA.395..226Z,1996ApJ...467..728M,2001ApJ...562..433E,2001ASPC..243..139K,2001MNRAS.323..785B,2006MNRAS.370..488B,Girichidis2012}, for instance, in the densest gas-rich areas where they can continuously accrete material competing with neighbour stars for this material, a scenario referred as competitive accretion
\citep{1982MNRAS.200..159L,1996ApJ...467..728M,1997MNRAS.285..201B}. 
Evidence of this scenario has been detected in embedded star clusters \citep[see e.g..][]{1996AJ....111.1964L,1997AJ....113.1733H,1998ApJ...492..540H,2006A&A...455..931B,2007AJ....134.1368C, 2013RAA....13..277E,2019A&A...629A.135D}. Mass segregation  has also been shown to develop dynamically \citep{2007ApJ...655L..45M,2009MNRAS.395.1449A,2011ApJ...732...16Y} and on short time scales \citep{2010MNRAS.407.1098A,2016MNRAS.459L.119P}, e.g., 
within $\sim$ 1 Myr. 
In some systems, dynamical processes are not fast enough to explain the high level of mass segregation observed ,
therefore some degree of mass primordial segregation is needed to explain such high concentration of massive stars \citep{1998MNRAS.295..691B,1998A&A...333..897R}. 

On the other hand, \citet{2015MNRAS.449.3381P} point out that stars formed by competitive
accretion rarely result in a segregated cluster. Motivated by the large theoretical and observational evidence that star clusters are formed in hierarchical distributions \citep{2010A&A...518L.106K, 1999AJ....118.1551W, 2000ApJ...545..327J, 2007ApJ...668.1042K, 2008MNRAS.389.1209S, 2009ApJS..184...18G, 2010A&A...518L..91D, 2011A&A...535A..77M, 2014MNRAS.438..639W}, \citet{Dom2017} found a level of mass segregation in the early stages of the embedded phase even starting with non-segregated substructured clusters, and also that a very high artificial level of mass segregation is not stable and it is quickly decreased by dynamical processes followed by a lower segregated state \citep[see also e.g.,][]{2009ApJ...700L..99A, 2010MNRAS.407.1098A}. If different levels of mass segregation affect the posterior evolution of clusters after gas expulsion is still an open question. 

Note though as a caveat to the hypothesis that star clusters form hierarchically from the mergers of smaller clusters: the projection of filamentary star formation within a part of a single molecular cloud, with well-separated and not merging embedded clusters spread along filaments, may appear, from the distant observer's point of view, as a sub-clustered cluster. \citet{BK2018} showed that the compactness and ages of observed very young clusters, such as the Orion Nebula Cluster, NGC3603 and R136, significantly constrain the hierarchical merging scenario.


\citet{Farias2015} and \citet{Farias2018} proposed two predictions to estimate the bound mass fraction remaining after violent gas expulsion for models which were representing embedded star clusters using as a first approach, equal-mass particles in order to study one parameter at time. Here, as a next step, we include the effects of particles having different masses following a typical initial stellar mass function (IMF) which add stronger two-body different mass interactions and mass segregation. The same sample is studied with stellar evolution (SEv) using winds and SNe. We test if the predictions proposed by \citet{Farias2015,Farias2018} are still valid using this expanded framework. 

The structure of the paper is a follows. In Section \ref{sec:methodinit} we describe our method and the initial conditions. In Section \ref{sec:sims} we explain our sample of simulations and in Section \ref{sec:results} we show our results. We end in Section \ref{sec:sumdisc} with our summary and conclusion.

\section{Method and initial conditions}
\label{sec:methodinit}
\subsection{Fractal distributions and initial mass function}
As in our previous work, we follow the setup described in \citet{Farias2015,Farias2018} and \citet{Dom2017}. Following the method described in \citet{2004A&A...413..929G}, we generate initially substructured distributions with a fractal dimension of $D = 1.6$, with a maximum radius of 1.5 pc and a total stellar mass of $500$ M$_\odot$.

\begin{figure}
    \centering
    \includegraphics[scale=0.69]{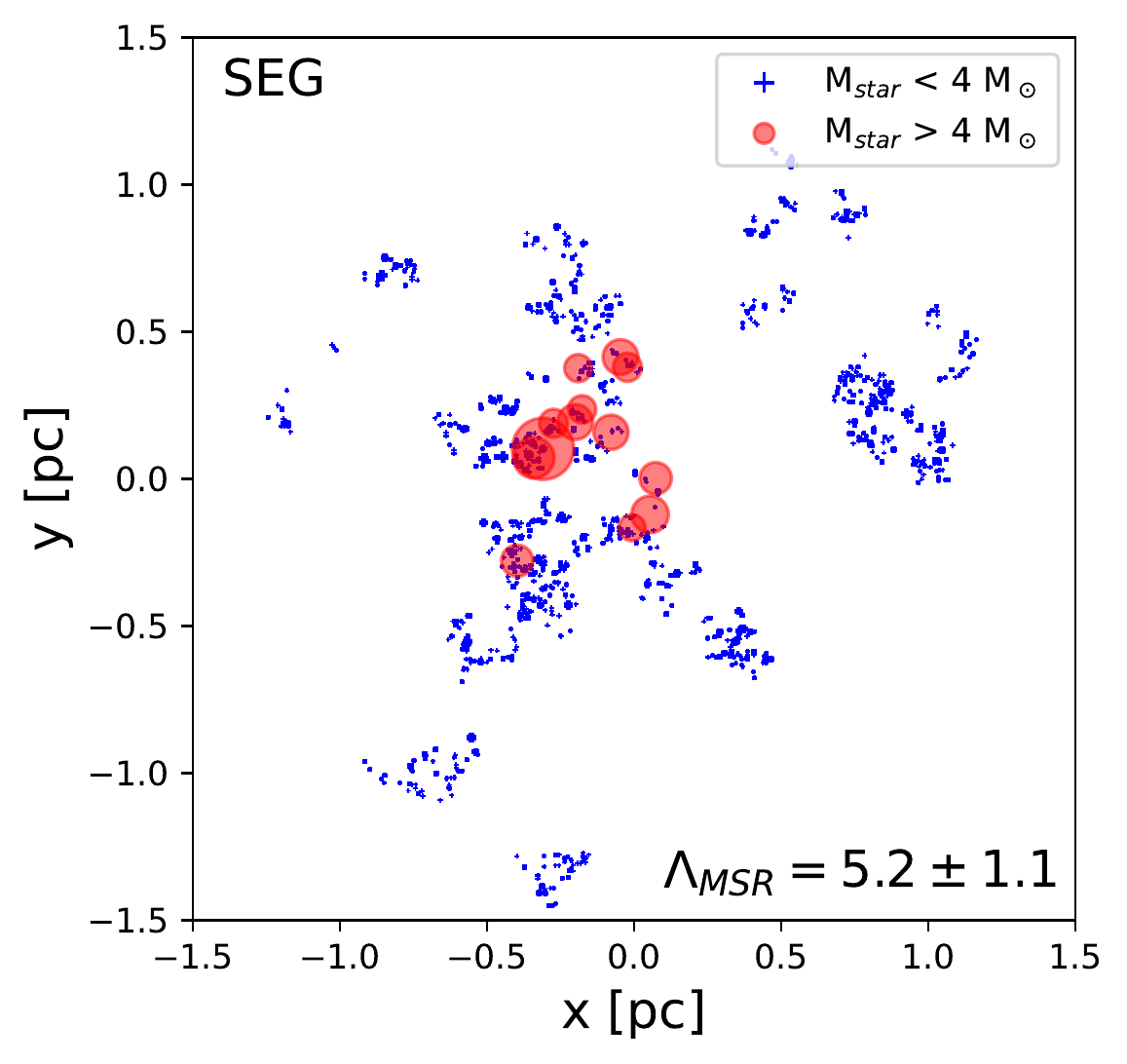}
     \includegraphics[scale=0.69]{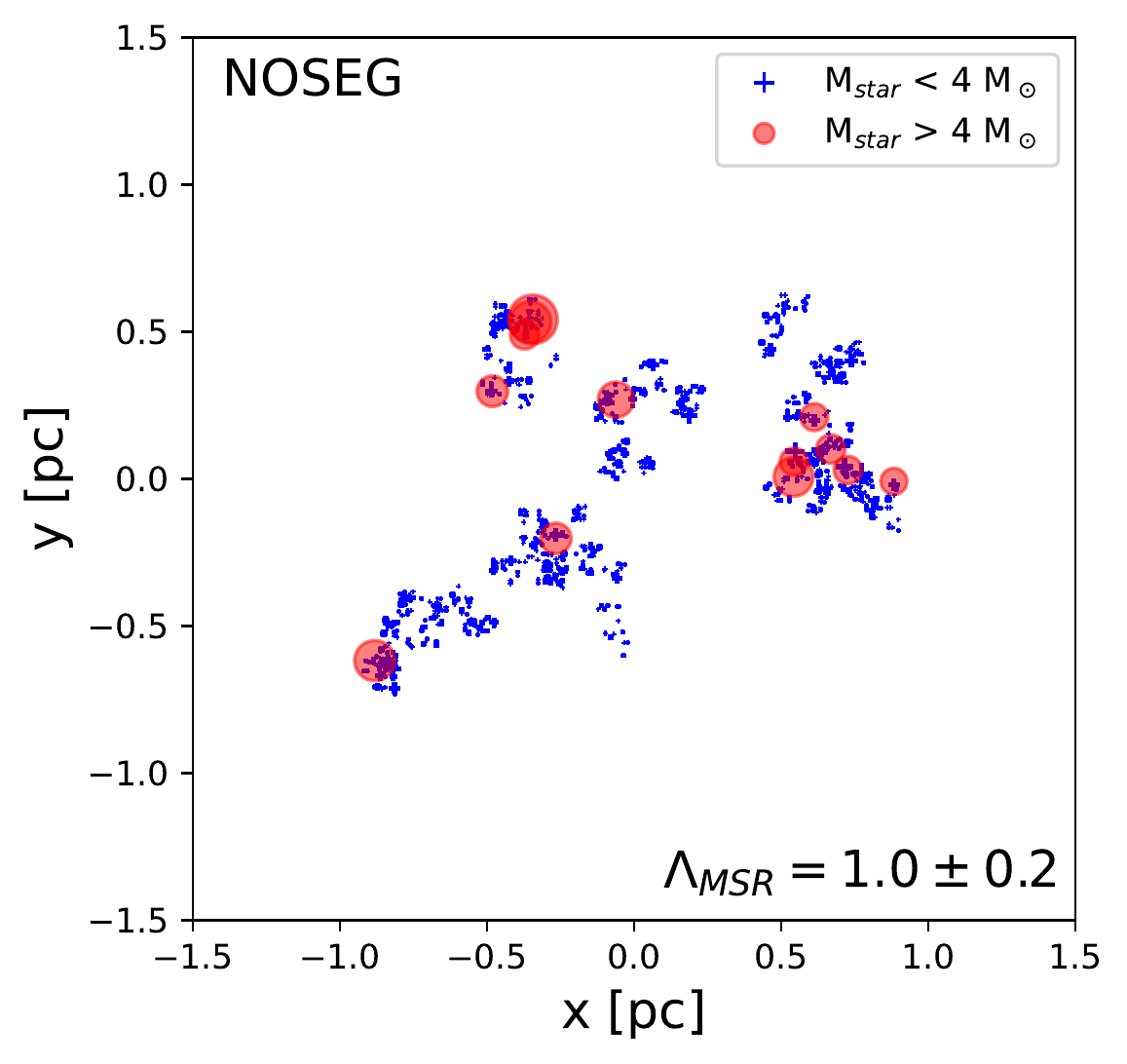}
    \caption{Two different fractal distributions. The top panel shows a cluster with mass segregation (SEG) and bottom panel a cluster non-segregated (NOSEG). Low mass stars ($M<$ 4 M$_\odot$) are represented with blue plus symbols (+) and massive stars ( $M\geq$ 4 M$_\odot$) with red circles. The sizes of the points are associated to the mass of the stars. The size of the massive stars is multiplied by 5 times for a better appreciation of their location. }
    \label{fig:Fractal}
\end{figure}

We assign individual stellar masses using the modified\footnote{We avoid the substellar mass range below 0.08 M$_\odot$ for brown dwarfs} initial mass function of \citet{2002Sci...295...82K} given by:
\begin{equation}
    N(M) \varpropto
    \begin{cases}
      M^{-1.30} & m_0 \leq M/\text{M}_\odot < m_1 \ \ \ \ \ \ \ \ \ \ \ \ \ \ \ \ \ \ \\
      M^{-2.30} & m_1 \leq M/\text{M}_\odot < m_2  \\
      M^{-2.35} & m_2 \leq M/\text{M}_\odot < m_3
    \end{cases}
\end{equation}
with $m_0=0.08$, $m_1=0.5$, $m_2=1.0$, $m_3=50$ M$_\odot$. 
Using this initial mass function, we obtain a total number of stars of $\sim 1000$, and a average stellar mass of 0.5 M$_\odot$. 

\subsection{Mass segregation using the \texorpdfstring{$\Lambda_\text{MSR}$} \ \ parameter}
\label{sec:lambdamsr}
The focus of this study is to examine if primordial levels of mass segregation influence the later evolution of star clusters. We define all stars with $M \geq 4$ M$_\odot$ as massive stars while the rest are considered low-mass stars. 
We quantify the different levels of mass segregation using the ``mass segregation ratio'' parameter ($\Lambda_\text{MSR}$) introduced by \citet{2009MNRAS.395.1449A}. $\Lambda_\text{MSR}$ is calculated by first finding  
the length of the shortest path joining the $N_\text{MST}$ most massive stars, i.e., the minimum spanning tree (MST) length, $l_{\text{massive}}$.  Secondly, the average  
MST length of $N_\text{massive}$ random stars $\langle l_{\text{norm}}\rangle$ is calculated 
with its associated standard deviation $\sigma_{\text{norm}}$. 
Finally, $\Lambda_\text{MSR}$ is defined as:
\begin{equation}
    \Lambda_\text{MSR}=\frac{\left<l_{\text{norm}}\right>}{l_{\text{massive}}} \pm \frac{\sigma_{\text{norm}}}{l_{\text{massive}}},
\end{equation}
where a value of  $\Lambda_\text{MSR} \sim $ 1 indicates no
mass segregation, i.e., low and high mass stars are uniformly distributed.  $\Lambda_\text{MSR} \gg $ 1 indicates strong mass segregation, i.e., massive stars are located close to each other. $\Lambda_\text{MSR} < $ 1 means
inverse mass segregation, i.e., high mass stars are more dispersed than low mass stars.
In this work, we explore different levels of mass segregation. These are achieved by locating the massive stars:
\begin{enumerate}
\item[i)] randomly in a radius $r> 0.5$ pc until finding $\Lambda_\text{MSR} \sim $ 1, i.e., a cluster without mass segregation which hereafter is refereed as NOSEG.
\item[ii)] We force all massive stars to be located in a radius $r< 0.5$ pc until obtaining  4 $< \Lambda_\text{MSR} < $ 5, i.e., primordial mass segregated clusters, hereafter refereed as SEG. 
\end{enumerate}

An example of two fractal distributions with different levels of mass segregation is shown in Fig.~\ref{fig:Fractal}. The top panel shows a  strongly mass segregated (SEG) fractal star cluster, for this case we have $\Lambda_\text{MSR} =  5.2 \pm 1.1$. The bottom panel shows a non-segregated (NOSEG) fractal star cluster with $\Lambda_\text{MSR} =  1.0 \pm 0.2$ where massive stars are spread along the distribution.
In both panels, blue plus symbols (+) represent low mass stars ($M<4$ M$_\odot$) and red circles represent massive stars ($M>4$ M$_\odot$). The sizes of the symbols are proportional to the mass of the stars, but for massive stars the sizes have been multiplied by 15 for better appreciation. 

\subsection{Background potential}
We use two different descriptions for the distribution of the background gas (BG). One, assuming the gas is centrally concentrated, represented by a
\citet{1911MNRAS..71..460P} sphere, with a density radial profile,
$\rho (r)$, described by
\begin{equation}
\rho (r)=\frac{3{M}_\text{Pl}}{4\pi R_\text{Pl}^3}\left( 1 + \frac{r^2}{R^2_\text{Pl}}\right)^{-\frac{5}{2}}
\end{equation}
with ${M}_\text{Pl}$ and $R_\text{Pl}$ the Plummer Mass and Plummer radius respectively, and $r$ being the distance to the centre of the cloud. 
The enclosed mass ${M} (r)$ within Plummer sphere is 
\begin{equation}
{M} (r) = {M}_\text{Pl}\frac{r^3}{R^3_\text{Pl}}\left( 1 + \frac{r^2}{R^2_\text{Pl}}\right)^{-\frac{3}{2}},
\end{equation}
which produces a BG potential $\phi (r)$ as follows:
\begin{equation}
    \phi (r)=-\frac{G{M}_\text{Pl}}{R_\text{Pl}}\left( 1 + \frac{r^2}{R^2_\text{Pl}}\right)^{-\frac{1}{2}},
\end{equation}
where G is the gravitational constant.

The second set of models assumes the background gas is uniformly distributed
within the cloud. In this case, the density profile is constant with a value:
\begin{equation}
\rho (r)=\frac{{3M}_\text{tot}}{4\pi r^3_c},~ r<r_c
\end{equation}
with ${M}_\text{tot}$ the total mass of the sphere and $r_c$ the radius of the sphere truncated to be 1.8 pc. 
 The enclosed ${M} (r)$ within a uniform sphere is described by: 
\begin{equation}
{M} (r) = \frac{{M}_\text{tot}}{r^3_c}r^3 ,~ r<r_c
\end{equation}
and its respective BG potential $\phi (r)$ inside and outside of the sphere as follows:
\begin{equation}
    \phi (r) = \frac{{GM}_\text{tot}}{2 r^3_c}\left(r^2-3r_c^2\right),~ r<r_c, 
\end{equation}
\begin{equation}
    \phi (r) = \frac{{GM}_\text{tot}}{r},~ r>r_c .
\end{equation}

The total mass for the background sphere of gas is chosen ensuring a global SFE $ = 0.2$ within a radius if 1.5 pc where the total mass in stars is $\sim500$ M$_\odot$. For the case of the Plummer sphere this is achieved by setting $M_\text{Pl}=3472$ M$_\odot$ and $R_\text{Pl}=1.0$ pc. For the uniform sphere case, $M_\text{tot}=3455$ M$_\odot$. We use these values in order to have a direct comparison with \citet{Farias2015,Farias2018} and \citet{Dom2017}, 
which are justified following a similar setup as in the classical picture of \citet{2007MNRAS.380.1589B} and observations \citep[see e.g.][]{2016AJ....151....5M}.

\begin{figure}
    \centering
    \includegraphics[width=\columnwidth]{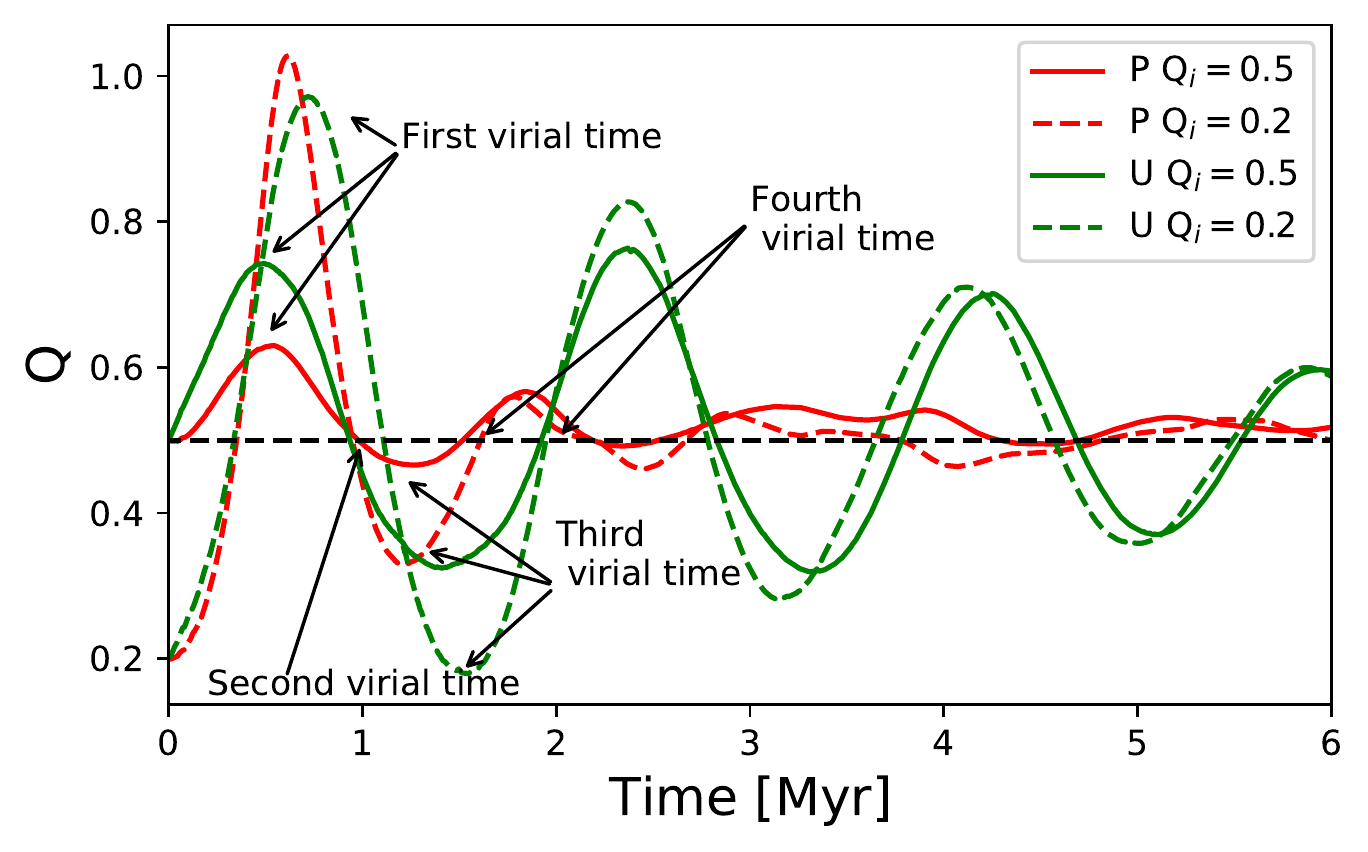}
    \caption{Example of $Q$ evolution in time with different initial virial states under different background gas distributions. Red line shows the evolution of $Q$ for a model under a Plummer (P) background gas profile and green line for a model with an uniform (U) distribution. 
    Solid and dashed lines shows models with virial ($Q_i=0.5$) and sub-virial ($Q_i=0.2)$ initial velocities.
     First, second, third and fourth virial times are pointed with arrows and they are the moments when the background gas is removed. Dashed black line shows the virial equilibrium value of $Q=0.5$.
     }
    \label{fig:tvsQ}
\end{figure}

\begin{table*}
    \centering
    \begin{tabular}{c|c|c|c|c|c|c|c}
         \hline
         Initial mass distribution& Initial virial ratio& Fractals &IMF&BG potential &Virial time &Number of simulations & Stellar evolution\\\hline
         SEG&0.5&10&10&Plummer/Uniform  & 1/2/3/4 &100/100/100/100&No\\
         SEG&0.2&10&10&Plummer/Uniform  & 1/2/3/4 &100/100/100/100&No\\
         SEG&0.5&10&10&Plummer/Uniform  & 1/2/3/4 &100/100/100/100&Yes\\
         SEG&0.2&10&10&Plummer/Uniform  & 1/2/3/4 &100/100/100/100&Yes\\[4pt]
         NOSEG&0.5&10&10&Plummer/Uniform  & 1/2/3/4 &100/100/100/100&No\\
         NOSEG&0.2&10&10&Plummer/Uniform  & 1/2/3/4 &100/100/100/100&No\\
         NOSEG&0.5&10&10&Plummer/Uniform  & 1/2/3/4 &100/100/100/100&Yes\\
         NOSEG&0.2&10&10&Plummer/Uniform  & 1/2/3/4 &100/100/100/100&Yes\\[4pt]
         EQUAL&0.5&100&0&Plummer/Uniform  & 1/2/3/4 &100/100/100/100&No\\
         EQUAL&0.2&100&0&Plummer/Uniform  & 1/2/3/4 &100/100/100/100&No
         \\\hline
         
    \end{tabular}
    \caption{Summary of initial conditions used for our study. The first column shows the initial stellar distribution, the second column shows the value of the initial virial ratio, the third and fourth columns indicate the number of different fractal distributions and IMF samples, respectively. 
    Background gas (BG) profile is provided in the fifth column. The sixth column indicates the different virial times when the gas is expelled, and the seventh column shows the number of realizations for each set. The eighth column shows cases where stellar evolution is included.}
    \label{tab:initcond}
\end{table*}

\subsection{Initial virial state}\label{sec:virialratio}
The virial ratio is defined as: 
\begin{equation}
Q = \frac{T}{|\Omega|},
\end{equation}
where $T$ and $\Omega$ are the total kinetic and potential energy of the system respectively\footnote{In latest literature $Q$ value is also found referred to $\alpha$ but we keep the symbol to be consistent with our line of papers.}.

In this work, we investigate two different initial dynamical states of star clusters, a sub-virial state represented by $Q=0.2$ and virial equilibrium state with $Q=0.5$. We note, however, that the latter does not represent a system in equilibrium, rather a system with velocities that match virial equilibrium. The fractal distributions used here are far from an equilibrium system, therefore, these systems will pursuit an equilibrium distribution. The rearrangement of stars and energy causes the measured virial ratio to oscillate around an equilibrium value as the clusters evolve. 
An example of the evolution of $Q$ with time under different conditions can be seen in Figure~\ref{fig:tvsQ} where the red line shows the evolution of one fractal cluster in a Plummer BG and the green line for the same fractal cluster but now under the influence of a uniform background gas distribution. Solid and dashed lines represent the initial states $Q_i=0.5$ and $Q_i=0.2$, respectively. The horizontal dashed black line represents the virial state $Q=0.5$. 
As expected sub-virial star clusters show a larger amplitude of the oscillation of $Q$ with time relative to the $Q_i=0.5$ case. Stars in a cluster with $Q_i=0.2$ tend to have orbits that fall through the center of potential of the system, reaching high velocities as they cross the potential minimum. While stars in systems with $Q_i =0.5$ tend to have more circular and stable orbits.

\subsection{Virial evolution and gas expulsion}
In Figure~\ref{fig:tvsQ} we point out different locations of the evolution of $Q$. At the selected points, we emulate rapid gas expulsion by removing the influence of the background gas. 
Different locations on the oscillation of the virial ratio are referred as virial time (VT). We call the first peak in the evolution of $Q$ "First virial time". At this point, star clusters are supervirial and we can obtain star clusters with pre-gas expulsion virial ratio of $Q_\text{f} > 0.5$. The exact values of $Q_\text{f}$ vary between the different models.
After this first maximum, the cluster passes through a state with $Q=0.5$, we term this point "Second virial time". All star clusters at this point have the same value of $Q_\text{f}$. 
Then, star clusters reach a first minimum of $Q$, the "Third virial time".
Here, star clusters have sub-virial velocities and therefore we can obtain star clusters with $Q_\text{f} < 0.5$ within a range of values.
Finally, star clusters reach $Q = 0.5$ again, after the first minimum, we call this point "Fourth virial time". 
By simplicity, we refer to them as VT $ = 1,2,3,4$, respectively. 
Note that these four points in the evolution of star clusters are different for each individual cluster. 
Therefore, for each set of initial conditions, we run 4 simulations removing the BG potential at these four different times. 

\subsection{Bound mass}
We refer as the bound mass \fbound to the fraction of stellar mass that is gravitationally bound ($M_\text{bound}$) relative to the initial stellar mass $M_\text{init}$:
\begin{equation}
\fbound=\frac{M_\text{bound}}{M_\text{init}}.
\end{equation}
We measure this value at different times in the evolution of the simulation. We compare the results with the two predictive models introduced by \citet{Farias2015} and \citet{Farias2018}. 
The first model is given by
\begin{equation}
    \fbound=\text{erf}\left(\sqrt{\frac{3}{2}\frac{\text{LSF}}{ Q_\text{f}}}\right)-\sqrt{\frac{6}{\pi}\frac{\text{LSF}}{ Q_\text{f}}}\text{exp}\left( -\frac{3}{2} \frac{\text{LSF}}{ Q_\text{f}}\right),
    \label{eq:model1}
    \end{equation}
where LSF is the local stellar fraction introduced by \citet{2011MNRAS.414.3036S} defined as the SFE measured within the stellar half-mass radius centered on one of the clumps and $Q_\text{f}$ is the pre-gas expulsion virial ratio, including the contribution of the BG to the potential felt by the stars. 
These two quantities are time dependent which contain more information about the stellar distribution at the time when they are measured. 

The second model neglects the contribution of the gas to the system. It estimates the amount of bound stellar mass using the virial ratio at the moment of gas expulsion ($Q_\text{a}$), assuming that all gas is expelled instantaneously. The bound stellar fraction is estimated as:
\begin{equation}
     \fbound=\text{erf}\left(\sqrt{\frac{3}{2}\frac{1}{ Q_\text{a}}}\right)-\sqrt{\frac{6}{\pi}\frac{1}{ Q_\text{a}}}\text{exp}\left( -\frac{3}{2} \frac{1}{ Q_\text{a}}\right).
     \label{eq:model2}
\end{equation}
In practice, this model simplifies the estimation of $\fbound$ as it only requires information from the stellar component.

These models were successfully tested in a scenario were all stars have the same mass. We also refer to these models as first and second prediction.
In this work, we test the reach of these models in a scenario where stars follow a realistic IMF and mass loss by stellar evolution is included.

\begin{figure*}
    \centering
    \includegraphics[width=\textwidth]{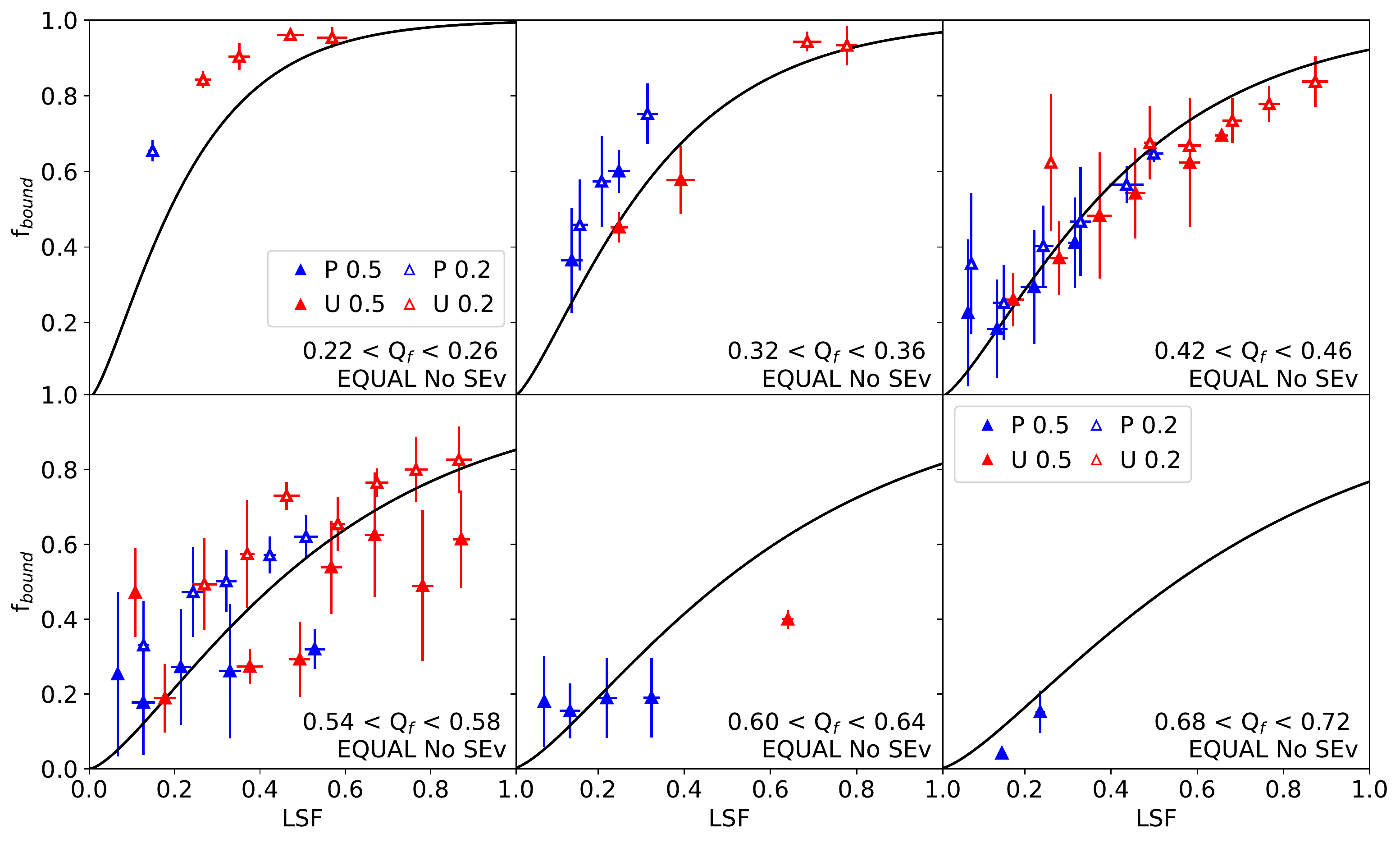}
    \caption{\fbound vs LSF for equal-mass particles simulations without stellar evolution at 16 Myr for VT = 1 (bottom row) and VT = 3 (top row). Blue and red triangles are simulations with Plummer (P) and Uniform (U) BG potential, respectively. The initial virial ratios $Q_\text{i}=0.5$ and $Q_\text{i}=0.2$ are represented by filled and empty symbols, respectively. The solid black line shows the predictive model introduced by \citet{Farias2015}, i.e., Eq.~\ref{eq:model1}, using the central value of $Q_\text{f}$ in each respective panel}
    \label{fig:LSFvsFBEQUAL}
\end{figure*}

\section{Set of simulations}
\label{sec:sims}
For the SEG sample, we create 10 fractal distributions and 10 different IMF samples associated with them. For each pair of positions and masses, we generate 10 different random assignments of the masses to the positions getting mass segregated clusters, which leads to a total number of 100 simulations. We double the number of simulations scaling the velocities of the particles in order to obtain embedded star clusters starting with $Q_i=0.5$ and with $Q_i=0.2$. For each $Q_i$ the sample is multiplied by four as we have four different VT where we remove the BG finalizing with 800 simulations. We add another sample of 800 simulations as we proceed in the same way to produce the NOSEG sample. We evolve the simulations for 16 Myr using the direct $N$-body code Nbody6++gpu \citep{2015MNRAS.450.4070W} which includes stellar mass loss from stellar evolution.
We double the 1600 SEG and NOSEG simulations running again the same sample, but this time with SEv activated. As we employ the latest version of the code used by \citet{Farias2015,Farias2018}, we also introduce a third sample with other 800 simulations, but this time based on equal-mass particles as a control method reproducing the results from our previous works. For the equal-mass particle sample, in order to have the same sample size, we use 100 different fractals and we proceed as before ending up with the same number of simulations. We do not use SEv for equal-mass particles simulations. Altogether, we perform a total number of 4000 simulations. The full sample is summarized in Table~\ref{tab:initcond}.

\begin{figure*}
    \centering
    \includegraphics[width=1.0\textwidth]{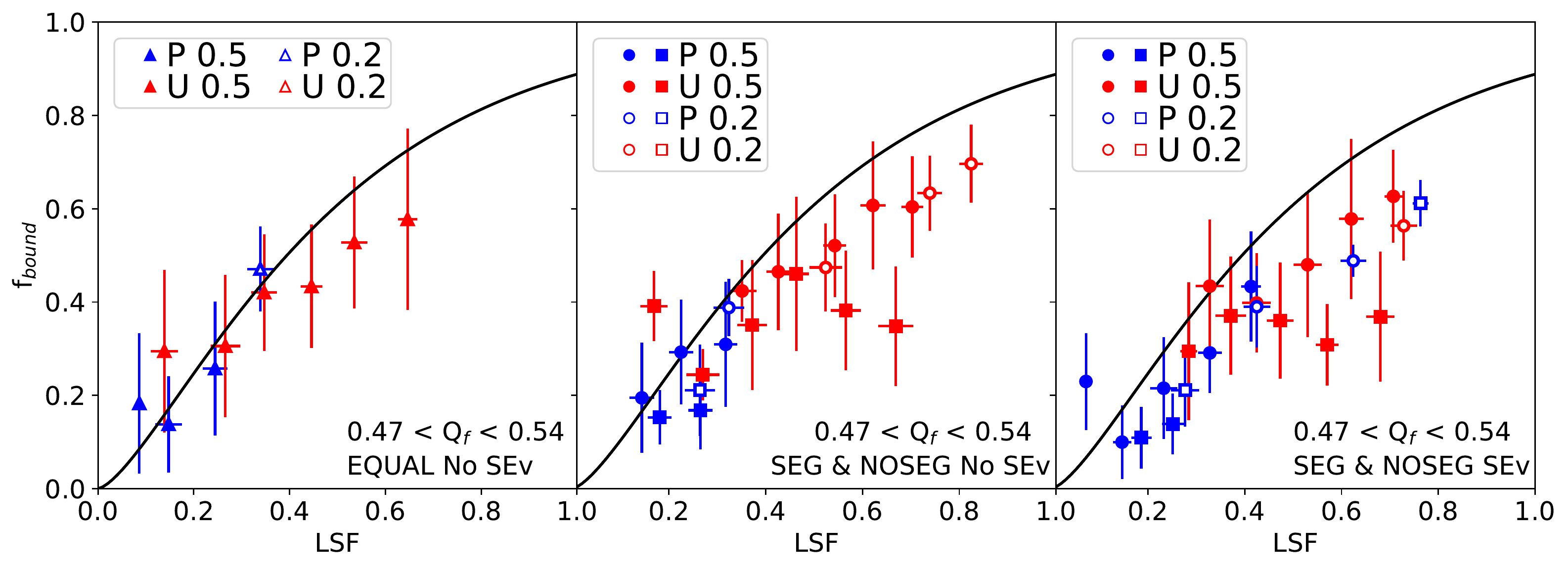}
     \caption{\fbound vs LSF only for $Q_\text{f}=0.5$ (only reached when VT = 2 and VT = 4). The left panel is for equal mass simulations and it uses the same colours and symbols as in Fig. \ref{fig:LSFvsFBEQUAL} where the black line is Eq. \ref{eq:model1}. The central panel shows results for simulations using IMF and non stellar evolution. SEG simulations are shown with circles and NOSEG simulations are shown with squares. The right panel is similar as the central one, but now for simulations using stellar evolution.}
     \label{fig:lsfqf05eqsegnosegnosevsev}
\end{figure*}

\begin{figure*}
    \centering
    \includegraphics[width=1.0\textwidth]{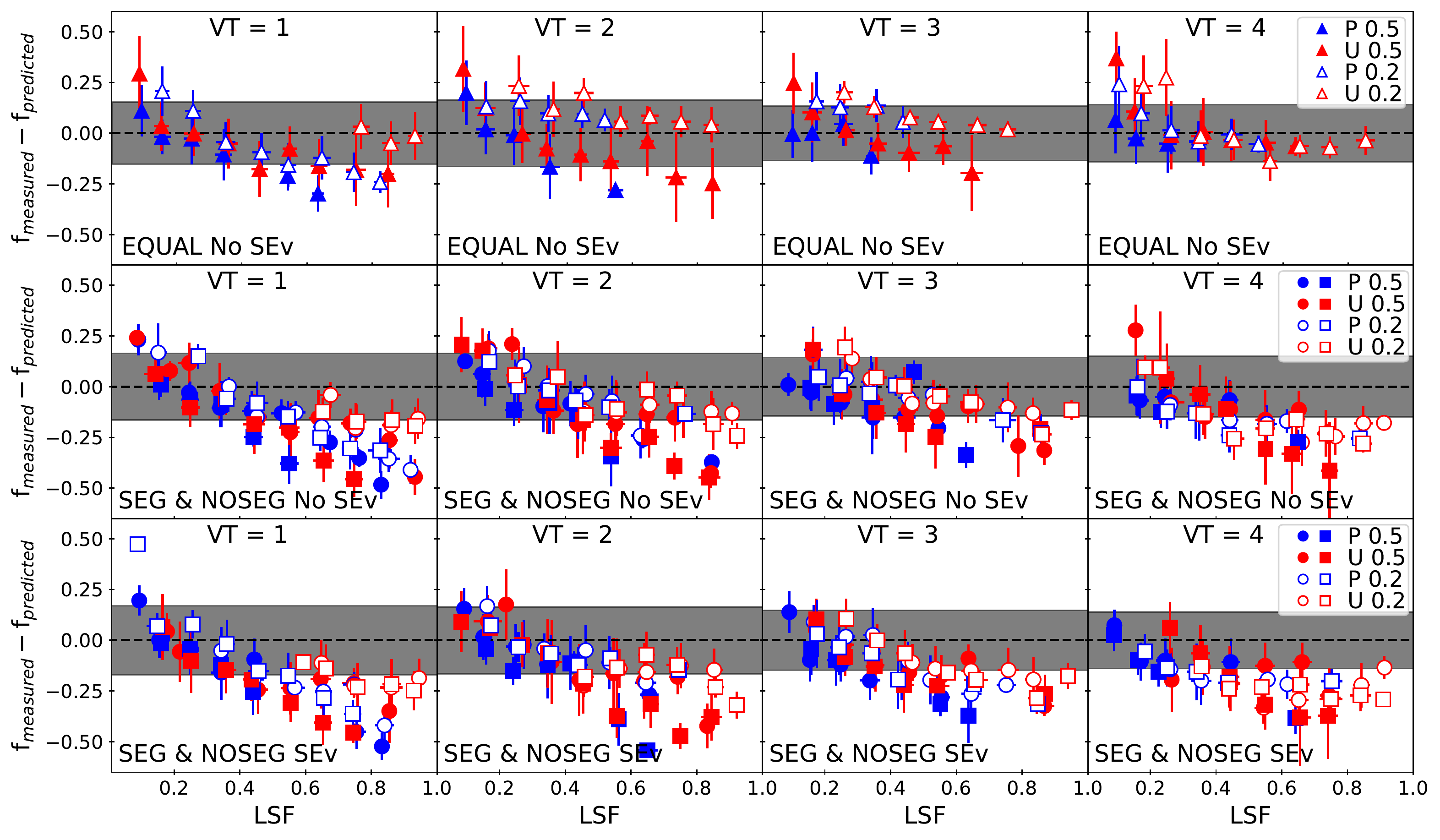}
    \caption{Average difference between the measured bound fraction ($f_\text{measured}$) and the predicted bound fraction ($f_\text{predicted}$) from  Eq.~\ref{eq:model1} vs LSF. The black dashed line represents a zero difference with the prediction. The panels are ordered from left to right by virial time (VT) and from top to bottom as in Fig. \ref{fig:lsfqf05eqsegnosegnosevsev} (now vertically) with the respective colour and symbols. The gray area represents the average 1 $\sigma$ error for all the result in each panel.}
    \label{fig:fbdifall16myrp1}
\end{figure*}

\begin{figure*}
    \centering
    \includegraphics[scale=0.55]{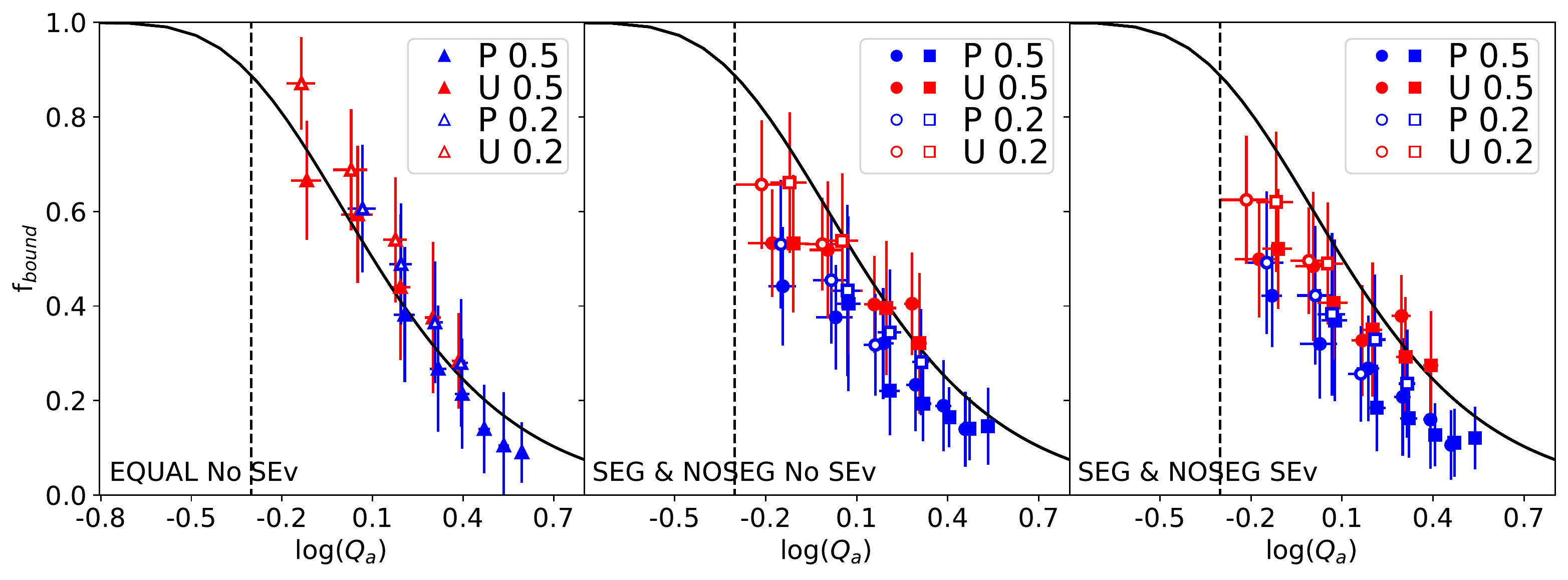}
    \caption{\fbound vs log($Q_\text{a}$). The order, colour and symbols are the same as Fig. \ref{fig:lsfqf05eqsegnosegnosevsev}. The solid black line shows the Eq.~\ref{eq:model2}. Dashed black line is the equilibrium value of $Q=0.5$ as a reference.}
    \label{fig:QavsFBEQUAL}
\end{figure*}

\begin{figure*}
    \centering
    \includegraphics[width=0.99\textwidth]{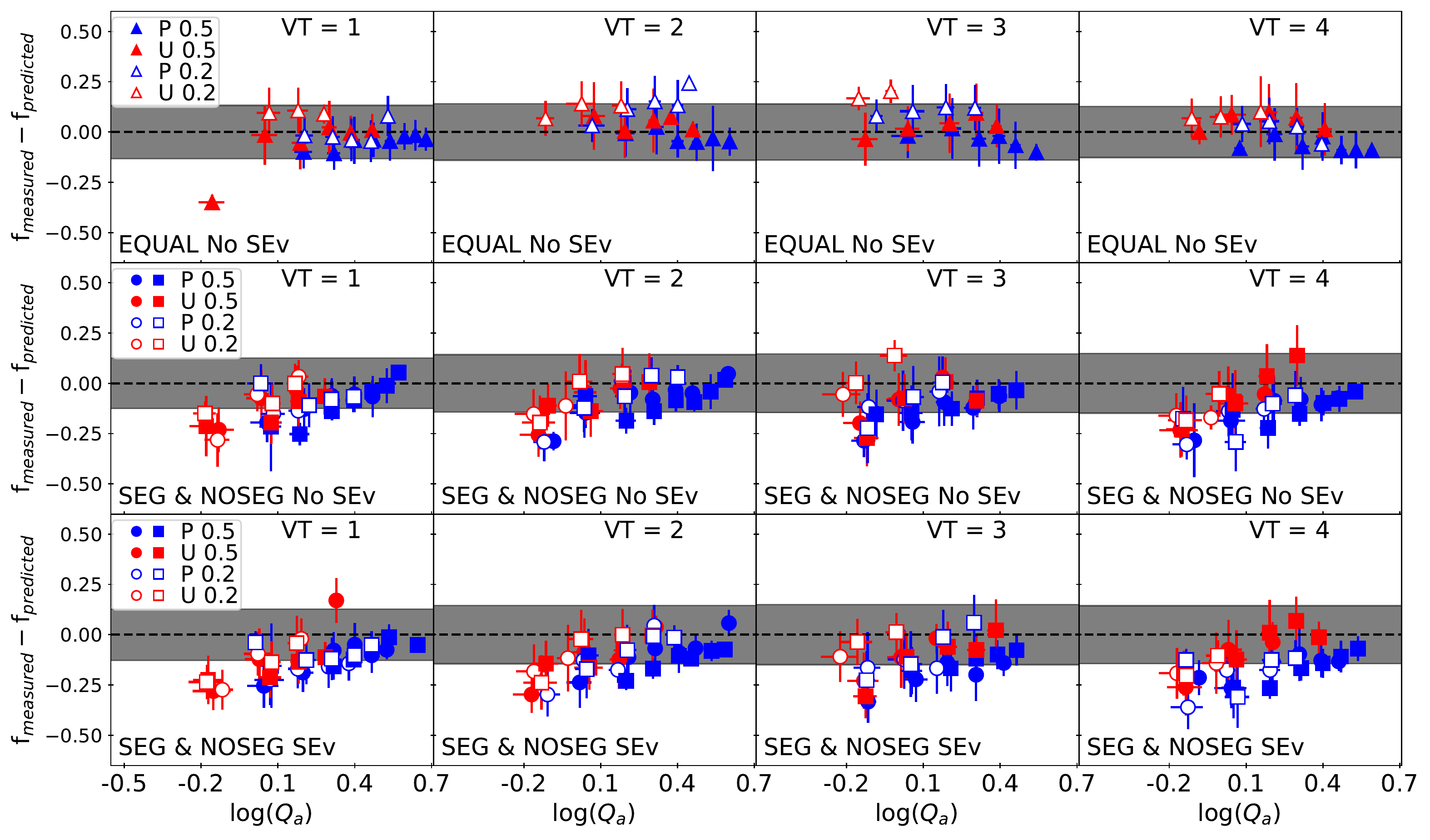}
    \caption{Average difference between the measured bound fraction ($f_\text{measured}$) and the predicted bound fraction ($f_\text{predicted}$) from  Eq.~\ref{eq:model2} vs $Q_\text{a}$. The order, colour and symbols are the same as in Fig. \ref{fig:fbdifall16myrp1}.}
    \label{fig:fbdifall16myrp2}
\end{figure*}

\begin{figure*}
    \centering
    \includegraphics[width=\textwidth]{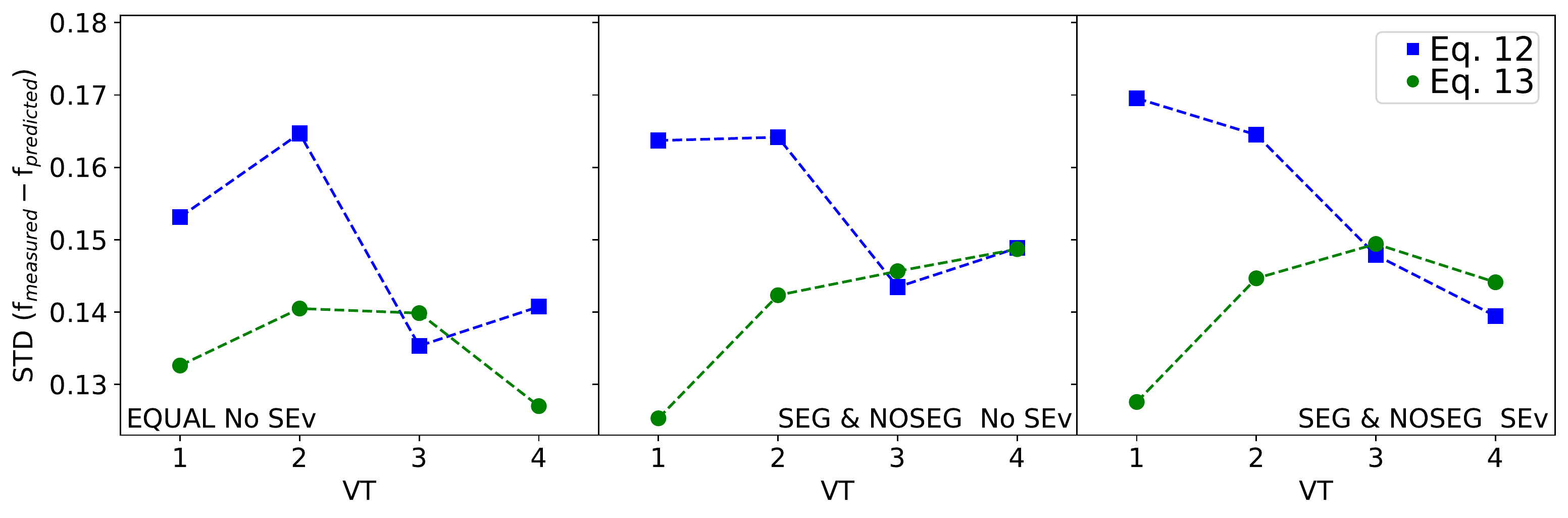}
    \caption{1 $\sigma$ error value (STD) of the difference between the predicted bound fraction ($f_\text{predicted}$) vs virial time. The squares with blue lines are for Eq.~\ref{eq:model1} and circles with green line are for Eq.~\ref{eq:model2}. The panel order is the same as in \ref{fig:lsfqf05eqsegnosegnosevsev}.} 
    \label{fig:stdfbdif16myr}
\end{figure*}

\begin{figure*}
    \centering
    \includegraphics[width=1.0\textwidth]{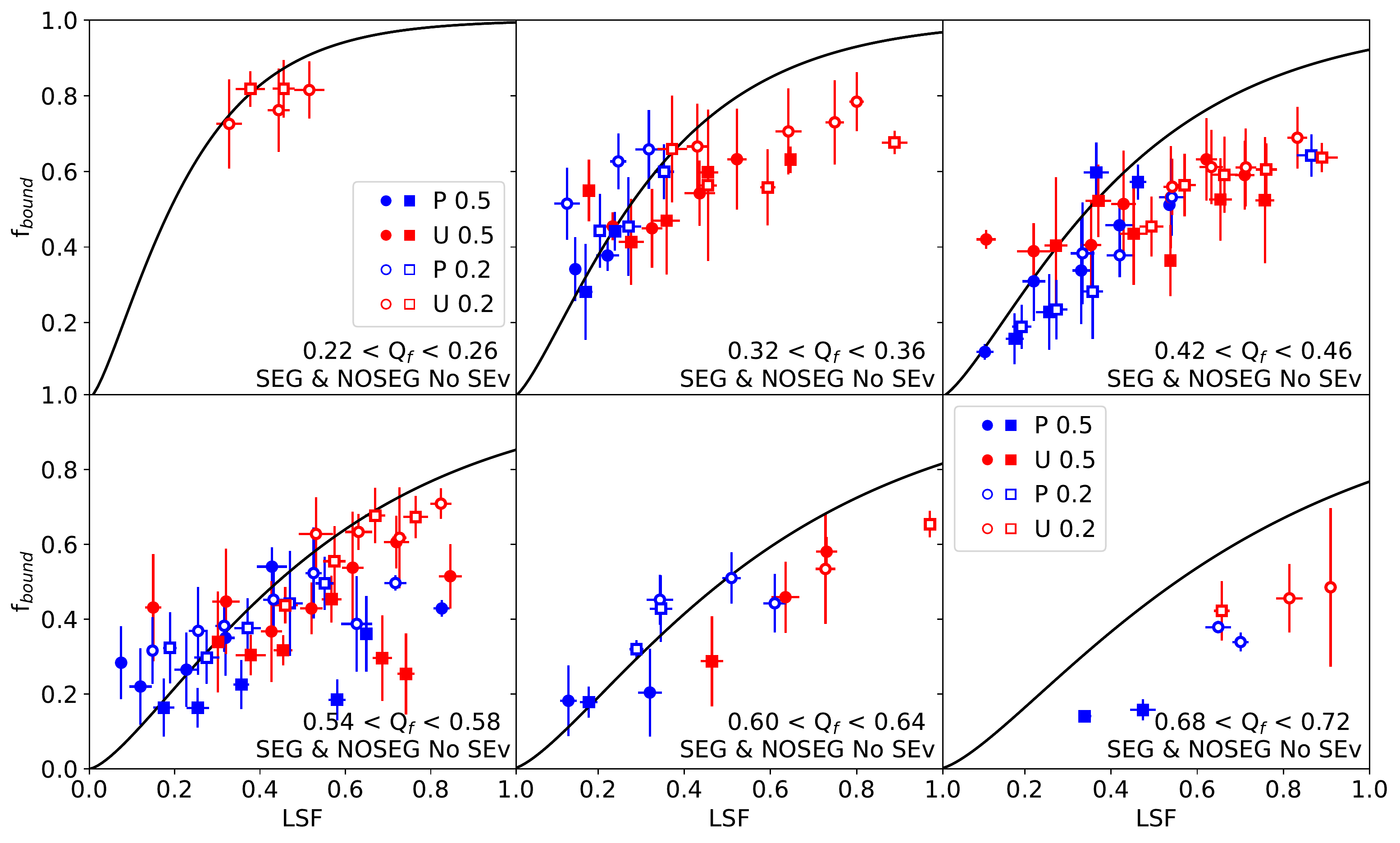}
    \caption{\fbound vs LSF for simulations with non stellar evolution at 16 Myr for VT = 1 (bottom row) and VT =3(top row). Colour and symbols are the same as in Fig. \ref{fig:lsfqf05eqsegnosegnosevsev} where the black line is Eq. \ref{eq:model1}.}
    \label{fig:LSFvsFBSEGNOSEGtog}
\end{figure*}

\section{Results}
\label{sec:results}
As in previous works, we are mostly interested in the fraction of stars that remain bound at a late stage, when any sign of initial structure is already lost, specifically we measure \fbound at 16 Myr which corresponds to $\sim 14.6$ initial crossing times. 
We reproduce the same plots as already shown in \citet{Farias2015} and \citet{Farias2018} for a direct comparison.
However, we notice that \fbound is not constant after gas expulsion, and therefore we also measure it at early stages in the evolution, i.e., we determine \fbound at the precise moment of gas expulsion (TEXP) and at times $t=4.8$, 6.4, 8, 9.6, 11.2, 12.8 and 13.4 Myr. In this way we can see how fast dynamical evaporation is affecting the surviving systems. Note that the exact value of TEXP is different for every cluster, since it is calculated based on their specific virial ratio evolution (see 
\S~\ref{sec:virialratio}).  
We use the same times, and $t = 0$ Myr, to observe what value of $\Lambda_\text{MSR}$ is achieved relative to the imposed initial conditions.

\subsection{Equal-mass simulations}
\label{sec:eqm}
The results of our first 800 simulations sample are used as the control sample and for comparison with the new parameter space introduced in this work.

We reproduce the same plot as shown in \citet{Farias2015} which contains only VT = 1 and VT = 3. In Fig.~\ref{fig:LSFvsFBEQUAL}, we show the resulting bound fractions for this set, measured at 16~Myr for star clusters with different $Q_\text{f}$ and background gas distributions. Black solid line shows Eq. \ref{eq:model1} using the central value of $Q_\text{f}$ described in each panel and the LSF value from $x$-axis. Blue triangles are simulations under a Plummer BG potential (P) and red triangles are simulations under a Uniform BG potential (U). Filled and empty symbols are representing the initial virial ratios $Q_\text{i}$ = 0.5 and $Q_\text{i}$ = 0.2, respectively.
The triangles have the average values for \fbound at 16 Myr for the simulations with values of LSF and $Q_\text{f}$ in the respective range. 
Note that for the cases where $Q_\text{f}$ is sub and super-virial, the exact value of $Q$ is not possible to fix, since each cluster reaches a different peak in $Q$ depending of the initial distribution of stars. Therefore, we can not fill each panel with the same quantity of points, with the most extreme values of $Q_\text{f}$ being the rarest.

We observe that the first prediction (Eq.~\ref{eq:model1}) is more accurate in top panels (star clusters with $Q_\text{f} < 0.5$) whereas there is no clear trend in the bottom panels. The reason of this is due to the high levels of substructure still present in the bottom panels. Gas expulsion at VT = 1 is very early and therefore substructure had not had time enough to be erased. The LSF value is sensible
to this effect as it needs to find the half-mass star radius centered in one of these sub-clusters. 
We also include the results with VT = 2 and VT = 4 in Fig. \ref{fig:lsfqf05eqsegnosegnosevsev} (left panel). In this case, early and late gas expulsion are mixed and a large dispersion is measured, but the prediction still matches the results in 1 $\sigma$ error range.
To see the effect of early and later gas expulsion more clearly, we show in Fig. \ref{fig:fbdifall16myrp1}, top panels, the average difference between the \fbound measured ($f_\text{measured}$) and the predicted  value from Eq.~\ref{eq:model1} ($f_\text{predicted}$) divided by VT from left to right. The gray area represents the 1 $\sigma$ error including all the cases, showing less dispersion for VT = 3 and VT = 4. In Fig. \ref{fig:stdfbdif16myr} (left panel) the blue squares represent the values from the gray zone in Fig. \ref{fig:fbdifall16myrp1} shown independently for a better appreciation. \citet{Farias2015} mostly explored gas expulsion times with VT > 3 when the initial substructure is mostly erased by dynamical processes \citep{2009ApJ...700L..99A,2014MNRAS.438..620P}. The prediction gets much closer to the results at VT $\geq$ 3 and it is expected to get even closer when gas expulsion happens later. 
Nevertheless, the moment of gas expulsion is kept as shown to make the study more realistic as it has been constrained that gas expulsion occurs very early for low mass clusters \citep{2020MNRAS.499..748D}.

\begin{figure*}
    \centering
    \includegraphics[scale=0.395]{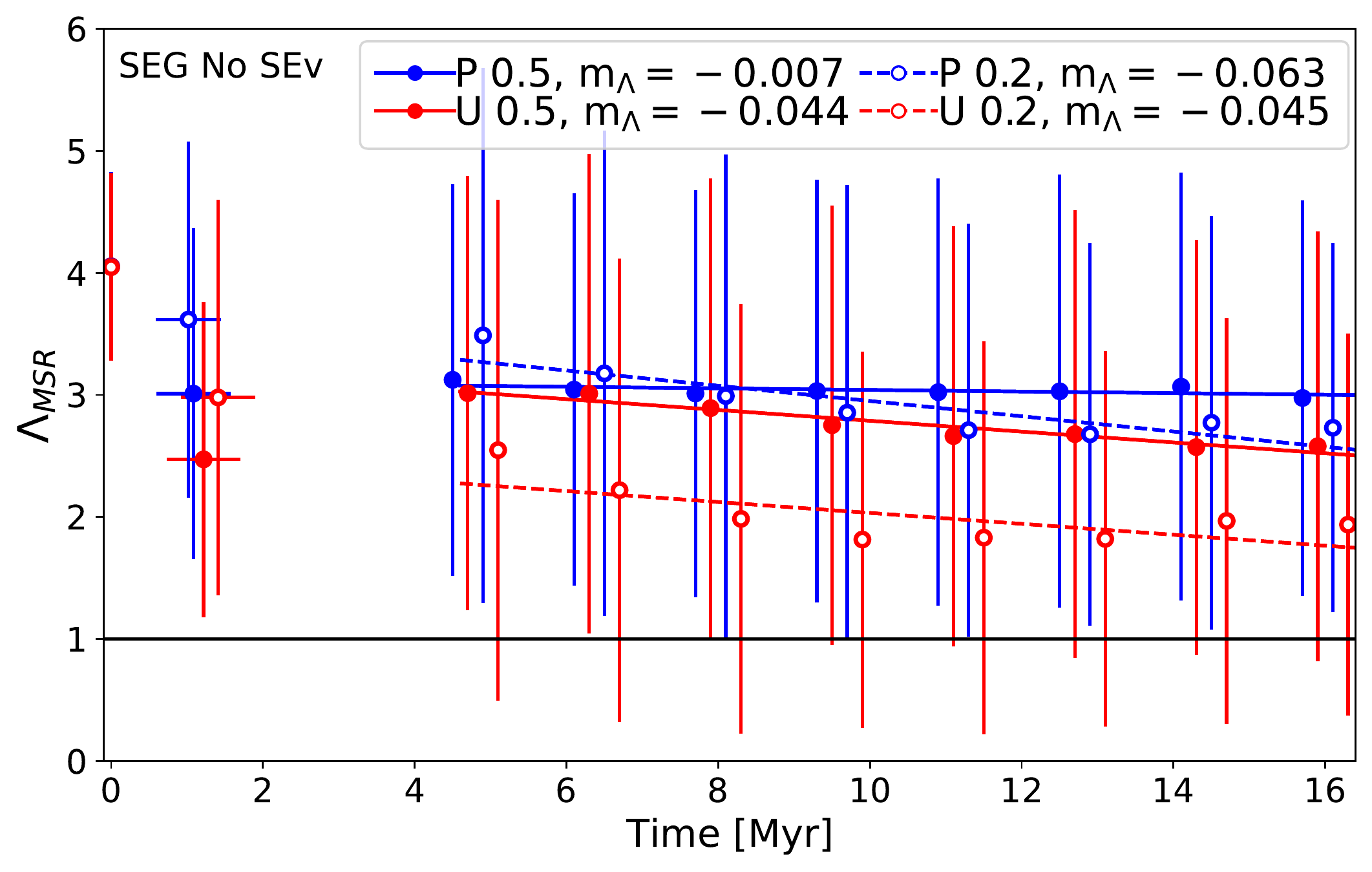}
    \includegraphics[scale=0.395]{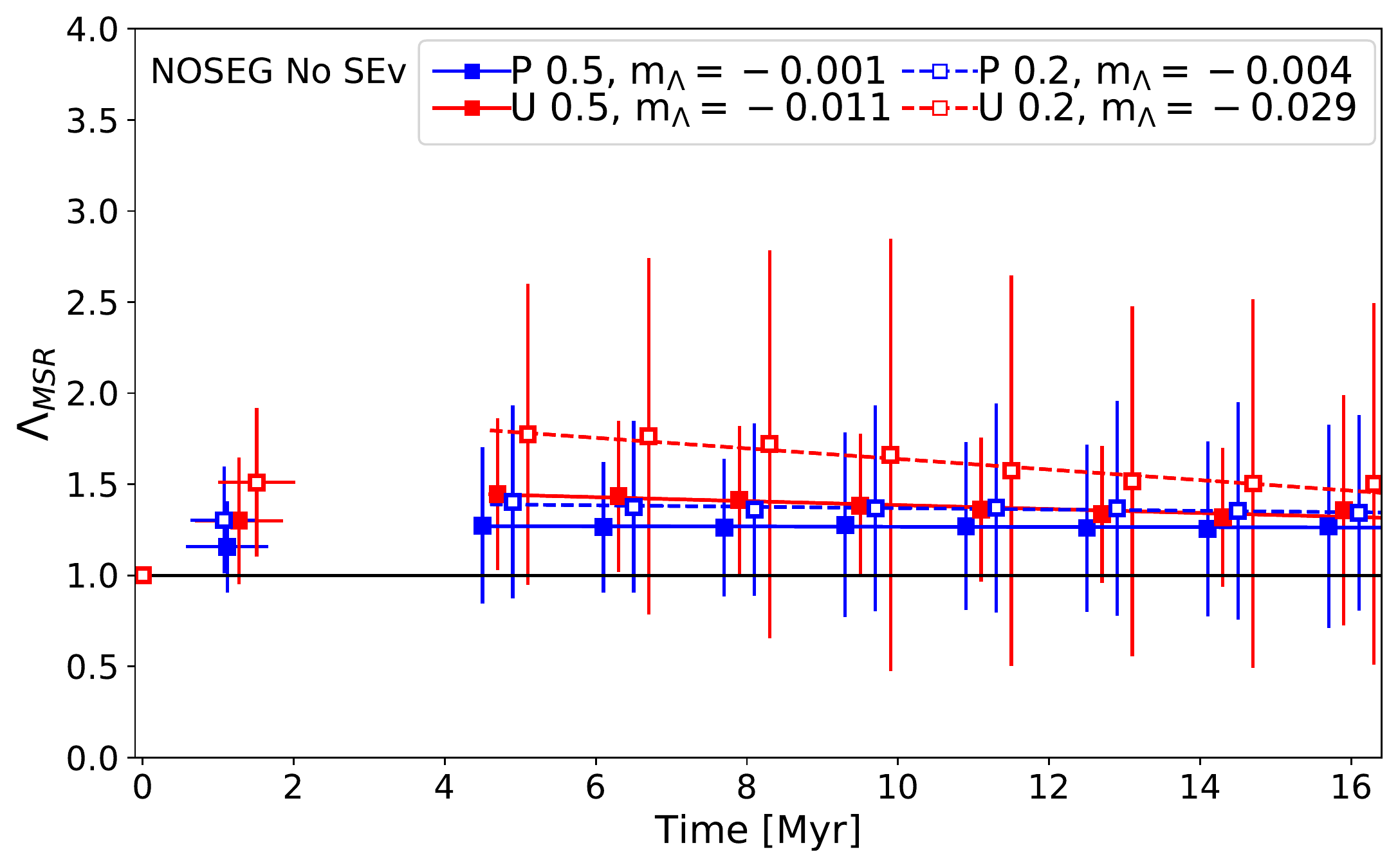}
    \includegraphics[scale=0.395]{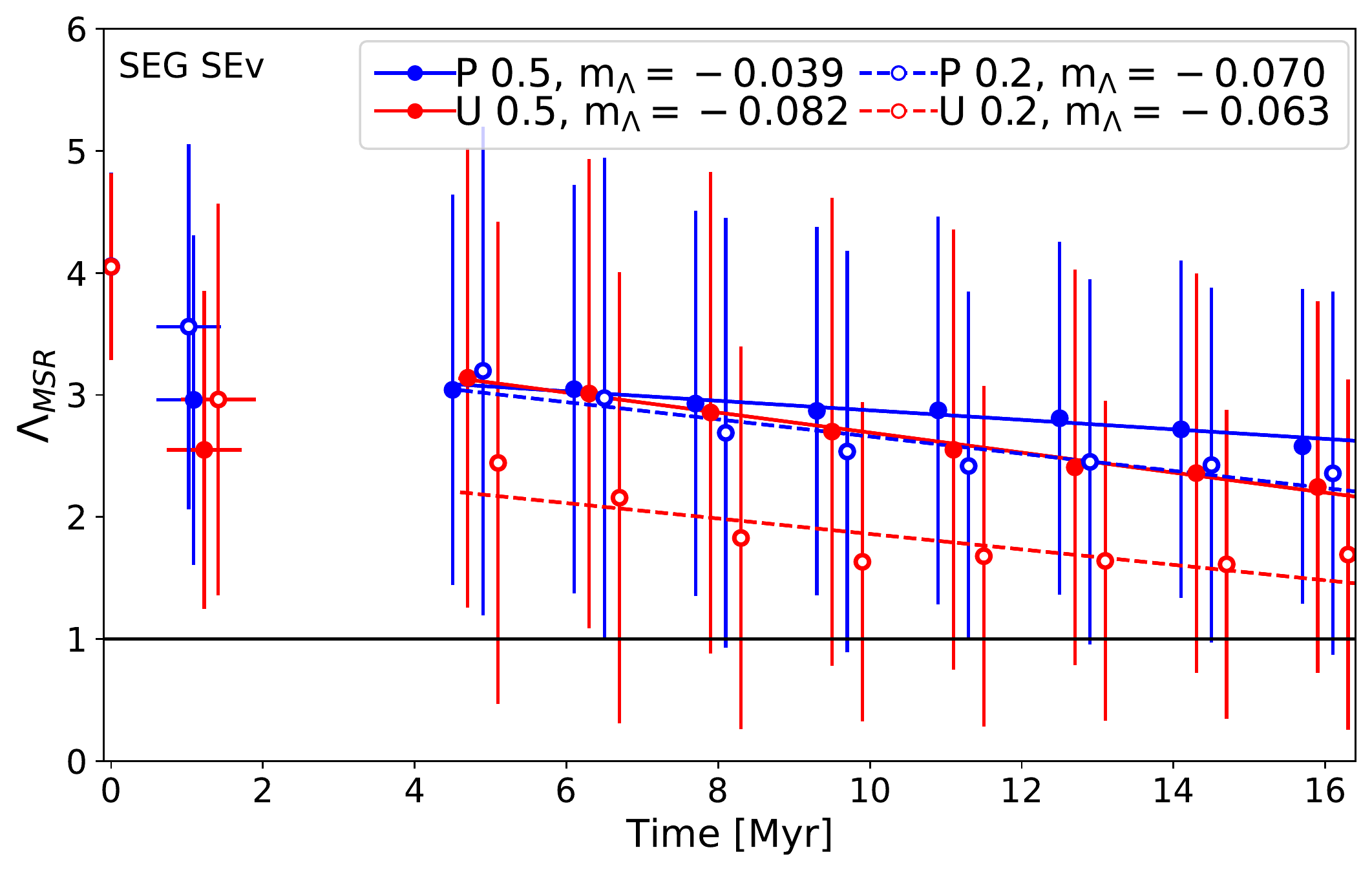}
    \includegraphics[scale=0.395]{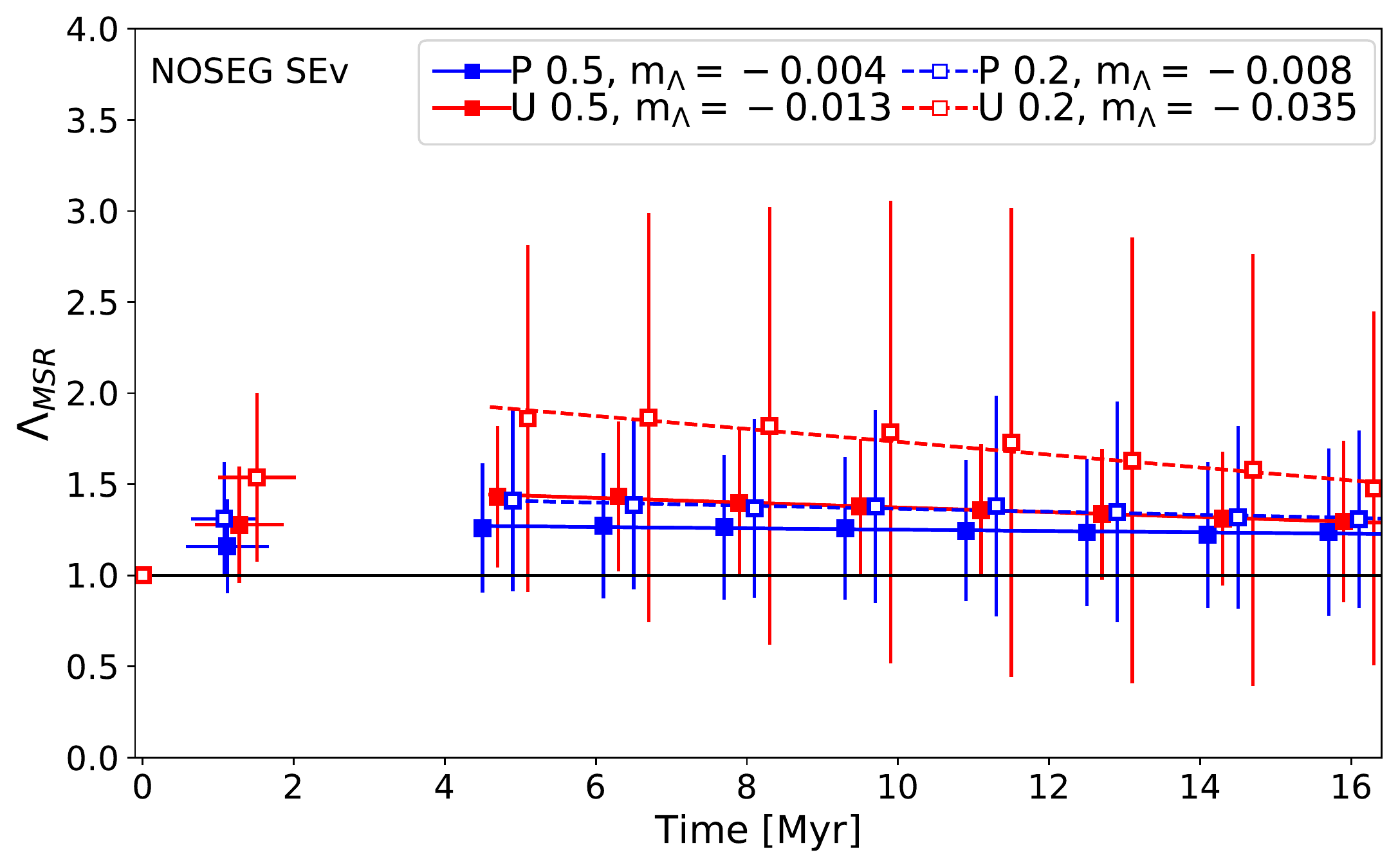}
    \caption{$\Lambda_\text{MSR}$ vs time. Colours and symbols are the same as in the previous figures.
    The solid black line is $\Lambda_\text{MSR}=1$ which means a star cluster not segregated. Simulations in top panels are with No-SEv and bottom panels with SEv. The lines show linear fit according to Eq.~\ref{eq:fitL} where the slopes $m_\Lambda$ are indicated respectively in the legends. 
    }
    \label{fig:tvsL}
\end{figure*}

In Fig. \ref{fig:QavsFBEQUAL} (left panel) we show the same models, again measured at 16 Myr, but as a function of the virial ratio right after gas expulsion, $Q_\text{a}$. Black solid line shows the prediction from Eq.~\ref{eq:model2}, i.e., using $Q_\text{a}$ as single parameter estimator. In this plot, all VT are included. Most of the clusters after gas expulsion become highly super-virial ($Q_\text{a} \gg 1$) because of the instant removal of the BG potential. Triangles represent the same initial conditions described before and the values are grouped in bins of $\Delta Q_\text{a}=0.5$. The model corresponding to Eq.~\ref{eq:model2} describes the results within the whole $Q_\text{a}$ range and it is not sensitive to the VT when the gas expulsion is measured. The latter can be seen in Fig.~\ref{fig:fbdifall16myrp1} (top panels), where a similar scatter is present when gas-expulsion happens at different VT with only a few exceptions falling outside the gray zone. The dispersion of the results is less for these results as 1 $\sigma$ error is smaller than before. Figure \ref{fig:stdfbdif16myr} (left panel) shows the width of the gray area for this prediction, with the green circles being smaller for three of the four VT. Therefore, the description provided by Eq.~\ref{eq:model2} is more suitable for our work since most of our simulations in this work expel the gas at very early times when high levels of substructure are still present.

\subsection{SEG-NOSEG simulations with no SEv}
\label{sec:segnosegnosev}
The results for SEG and NOSEG simulations with No-SEv at 16 Myr for VT = 1 and VT = 3 are shown in Fig. \ref{fig:LSFvsFBSEGNOSEGtog}. The symbols are the same as before. For these cases, circles indicate simulations starting with SEG and squares for simulations starting with NOSEG.

We observe that in most cases Eq. \ref{eq:model1} over-estimates \fbound, especially at higher values of LSF. We also include the results when $Q_\text{f}=0.5$ (VT = 2 and VT = 4) in Fig. \ref{fig:lsfqf05eqsegnosegnosevsev} (central panel), where we observe the same behaviour. 

We test Eq.~\ref{eq:model2} in Fig.~\ref{fig:QavsFBEQUAL} with the same symbols as before, where all VT are included. Again, at 16 Myr, most of the simulations have lower values of \fbound than expected. SEG and NOSEG simulations show the same behaviour, with both analytical models over estimating the bound fraction. When using Eq. \ref{eq:model1} many dots are outside the one-sigma error bars, while Eq. \ref{eq:model2} does a better job with estimations mostly within error bars.

The question of how early and late gas expulsion influence the accuracy of the prediction is addressed in Fig.~\ref{fig:fbdifall16myrp1} and  Fig. \ref{fig:fbdifall16myrp2} (second row) for the first prediction and second prediction, respectively. For Eq. \ref{eq:model1}, we observe that independent of the VT, the results are mostly out of the gray zone especially for higher LSF. For Eq. \ref{eq:model2}, more results are falling inside 1 $\sigma$ error zone with exceptions for low $Q_a$. This behaviour is better appreciable in Fig. \ref{fig:stdfbdif16myr}, where the dispersion for the results of the first prediction (blue squares) compared to the second prediction (green circles) is larger for VT $\leq 2$ and in the same range for VT $\geq 3$.
Besides the \textit{individual results}, in most of the cases, \fbound is found below the predictions.

\subsection{SEG-NOSEG simulations with SEv}
\label{sec:segnosegsev}
\begin{figure*}
    \centering
    \includegraphics[width=\textwidth]{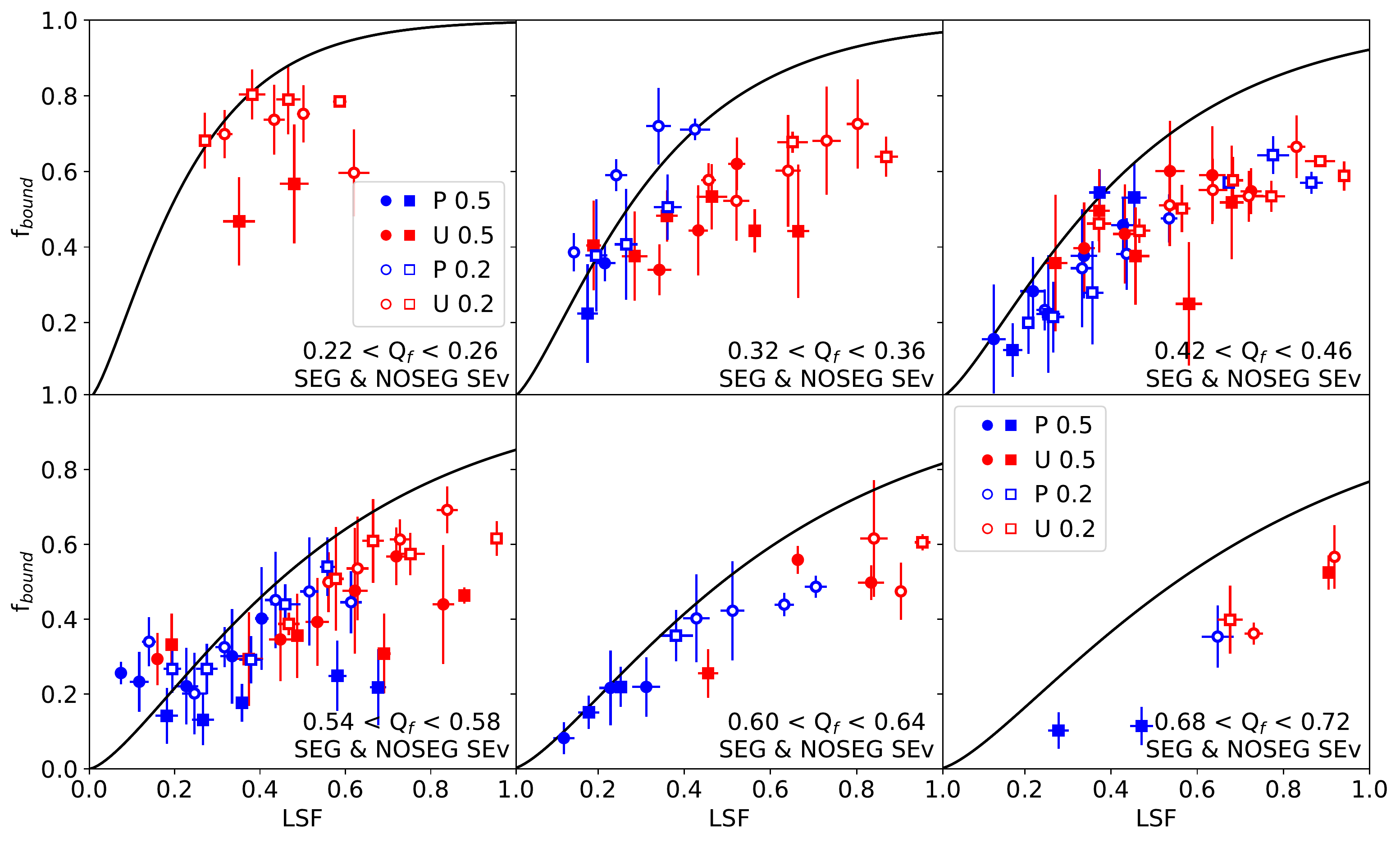}
    \caption{\fbound vs LSF for simulations with non stellar evolution at 16 Myr for VT = 1 (bottom row) and VT = 3 (top row). Colour and symbols are the same as in Fig. \ref{fig:lsfqf05eqsegnosegnosevsev}.}
    \label{fig:LSFvsFBSEGNOSEGtogsev}
\end{figure*}

The results for SEG and NOSEG simulations with SEv at 16 Myr for VT = 1 and VT = 3 are shown in Fig.~\ref{fig:LSFvsFBSEGNOSEGtogsev}. The symbols are the same as before. The values of LSFs are not expected to be identical for all pairs of simulations started with SEv and No-SEv due to small changes in the orbit calculations done in a $N$-body simulation. 
We observe as before that the prediction Eq.~\ref{eq:model1} over estimates \fbound. The number of simulations far from the curve is higher for this sample and the same is observed for $Q_\text{f}=0.5$ (VT = 2,4) in Fig. \ref{fig:lsfqf05eqsegnosegnosevsev} (right panel).

The prediction from Eq. \ref{eq:model2} is shown in Fig. \ref{fig:QavsFBEQUAL} with the same symbols as before. The values of \fbound at the end of most simulations are even smaller than when we do not use SEv. 

As in the previous cases, we test early and late gas expulsion in Fig.~\ref{fig:fbdifall16myrp1} and  Fig. \ref{fig:fbdifall16myrp2} (bottom panels). The same description previously mentioned for the simulations without SEv is applicable for these results but now with a larger number of simulations outside of the gray area. In Fig. \ref{fig:stdfbdif16myr} we observe a small improvement in the dispersion of the results when VT = 4, if we compare with its pair in the central panel.

\begin{figure*}
    \centering
    \includegraphics[scale=0.395]{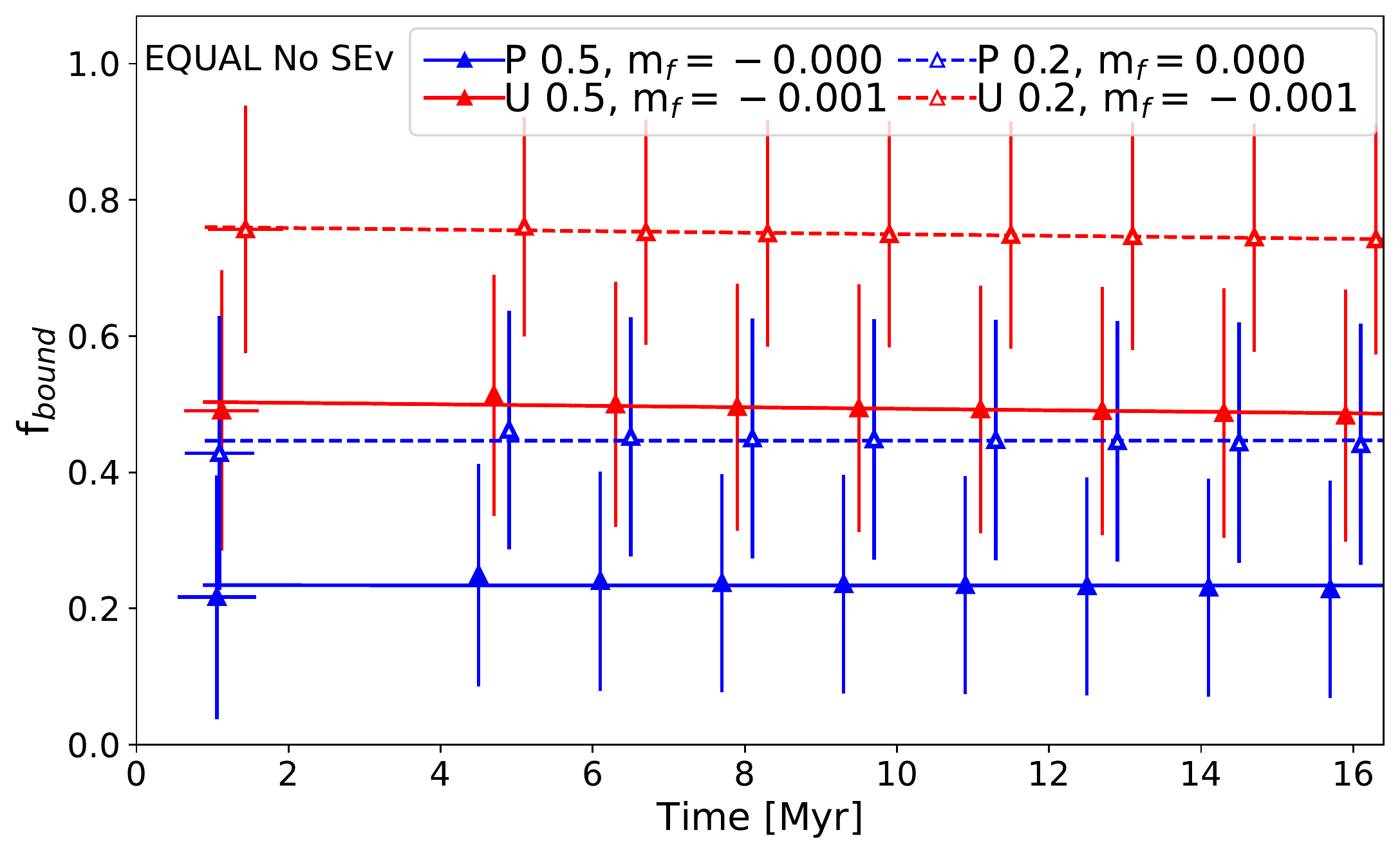}
    
    \includegraphics[scale=0.395]{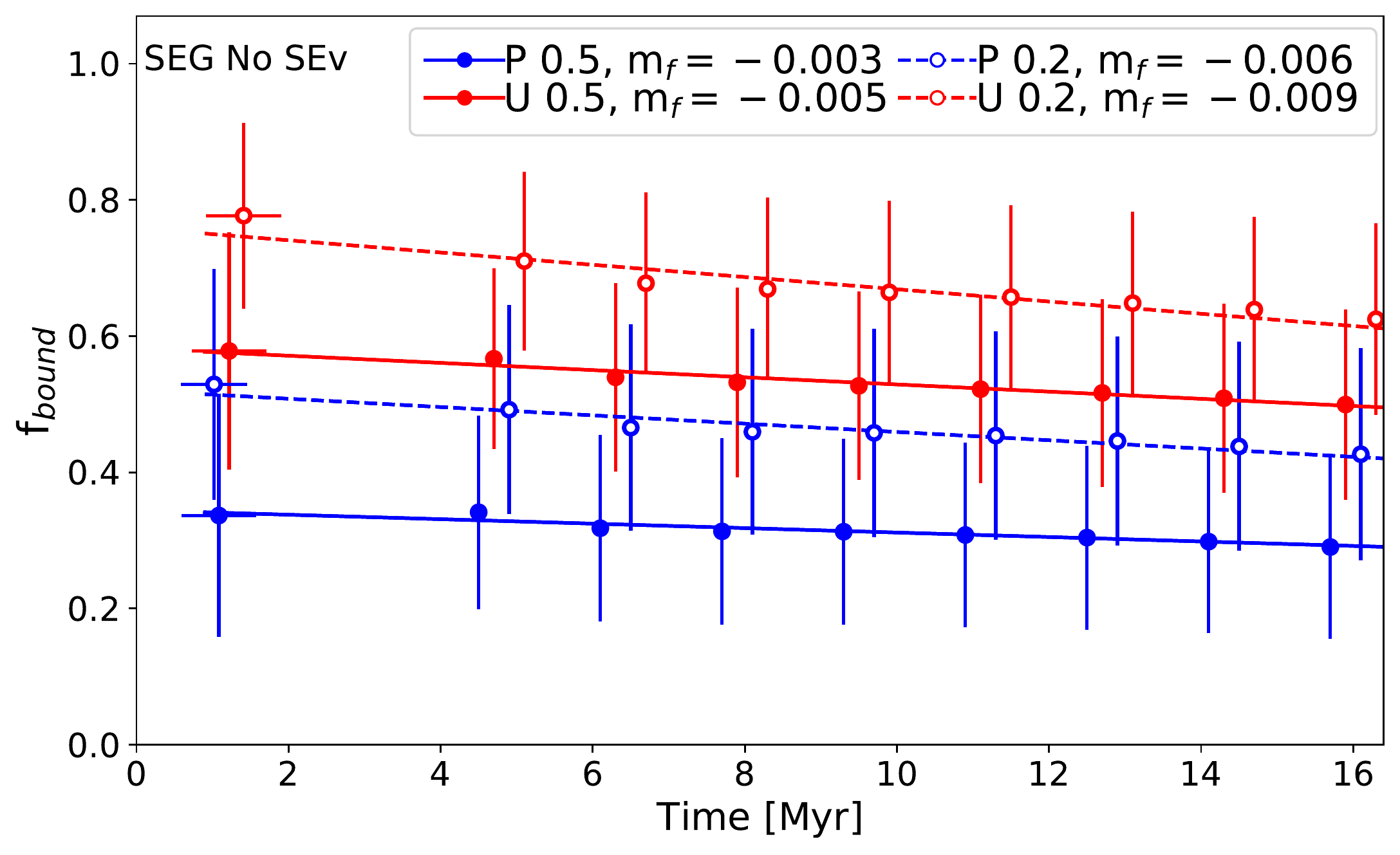} \includegraphics[scale=0.395]{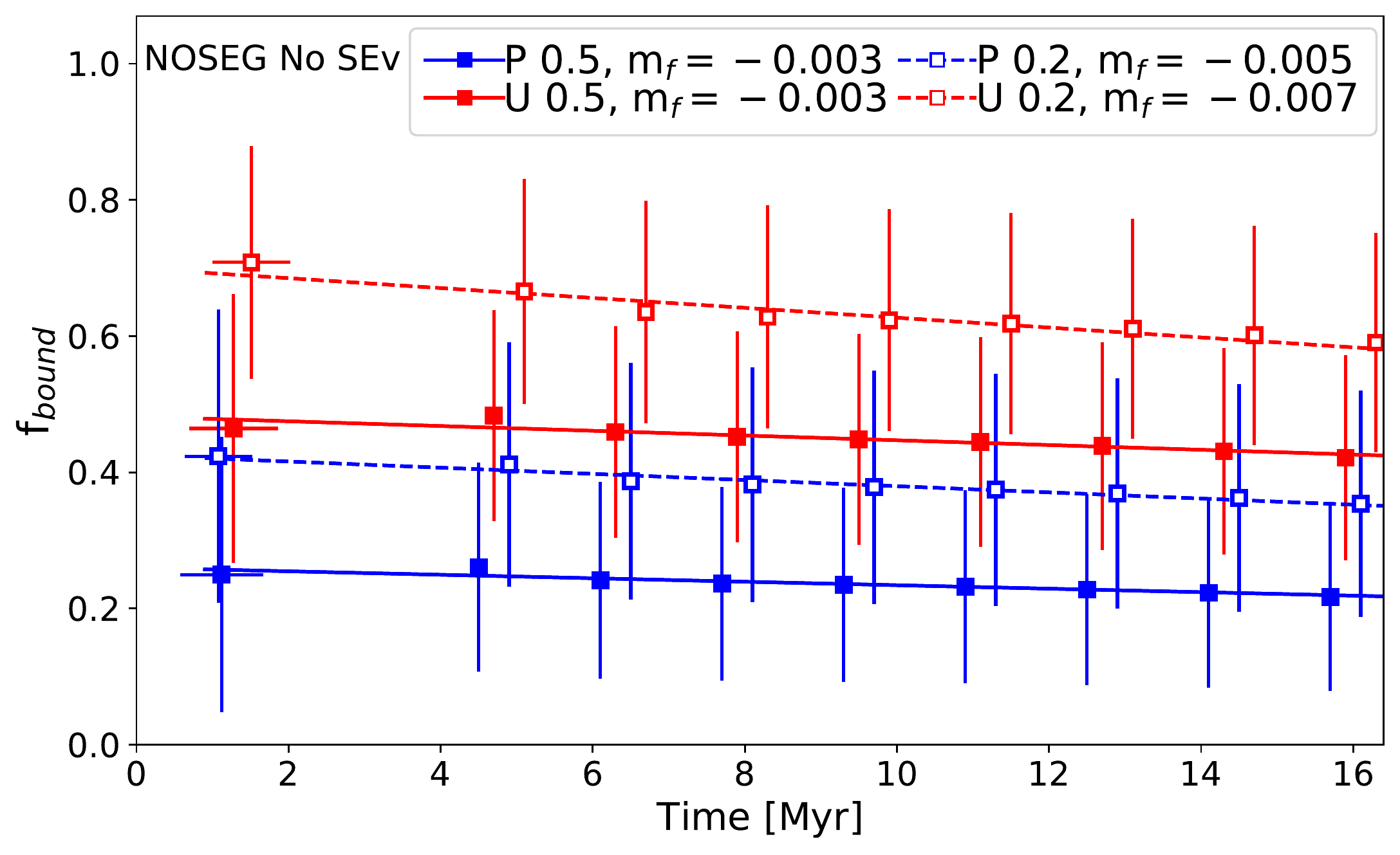}
    
    \includegraphics[scale=0.395]{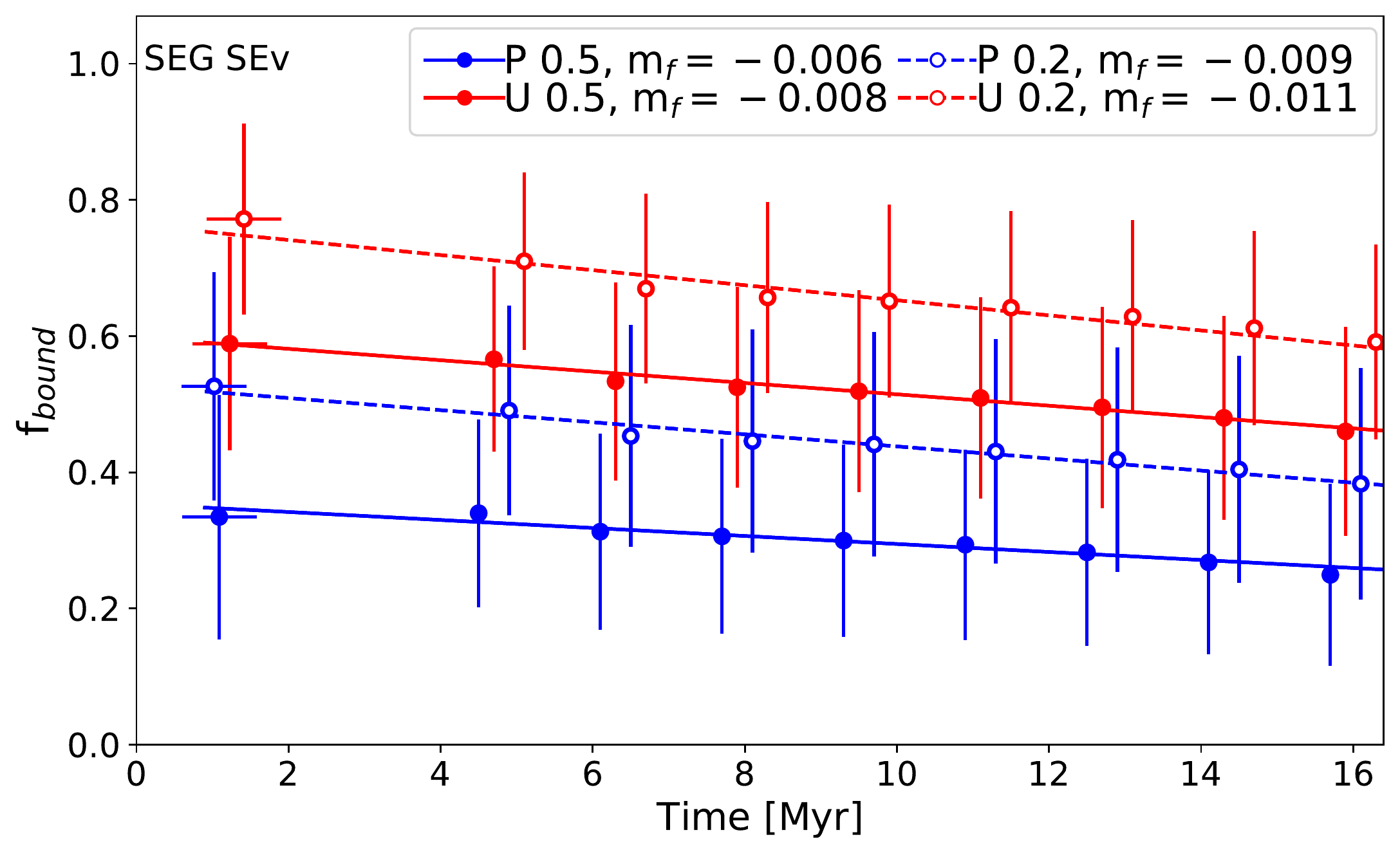} \includegraphics[scale=0.395]{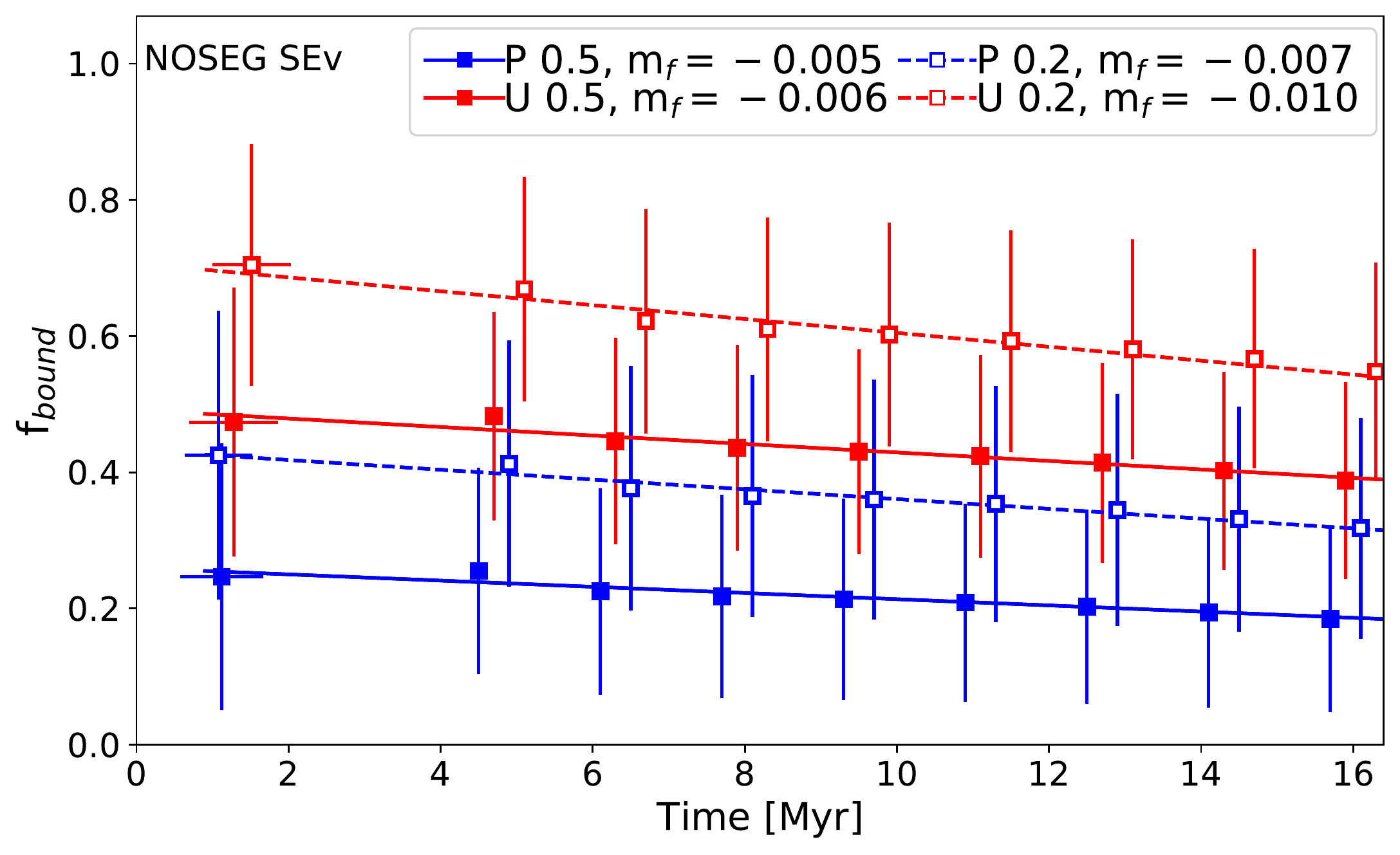}
    \caption{\fbound evolution in time. Colours and symbols are the same as in previous figures. 
    The lines show linear fit according to Eq.~\ref{eq:fitevap} where the slopes
    $m_f$ are indicated respectively in the legends. 
    }\label{fig:tvsFB}
\end{figure*}

\begin{figure*}
    \centering
    \includegraphics[width=0.99\textwidth]{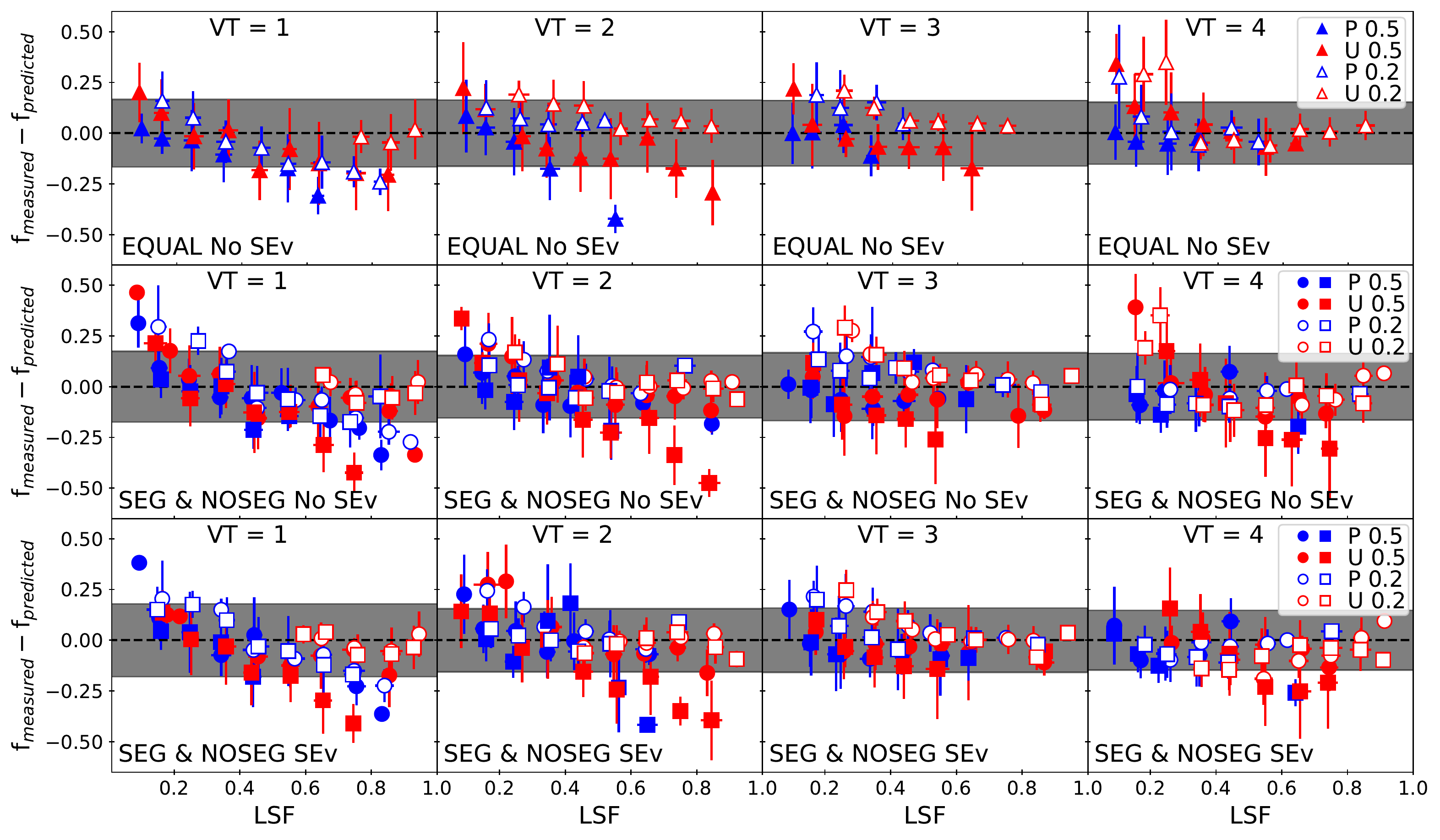}
    \caption{Same as Fig. \ref{fig:fbdifall16myrp1} but for the moment of the gas expulsion.}
    \label{fig:fbdifallTEXPp1}
\end{figure*}

\begin{figure*}
    \centering
    \includegraphics[width=0.99\textwidth]{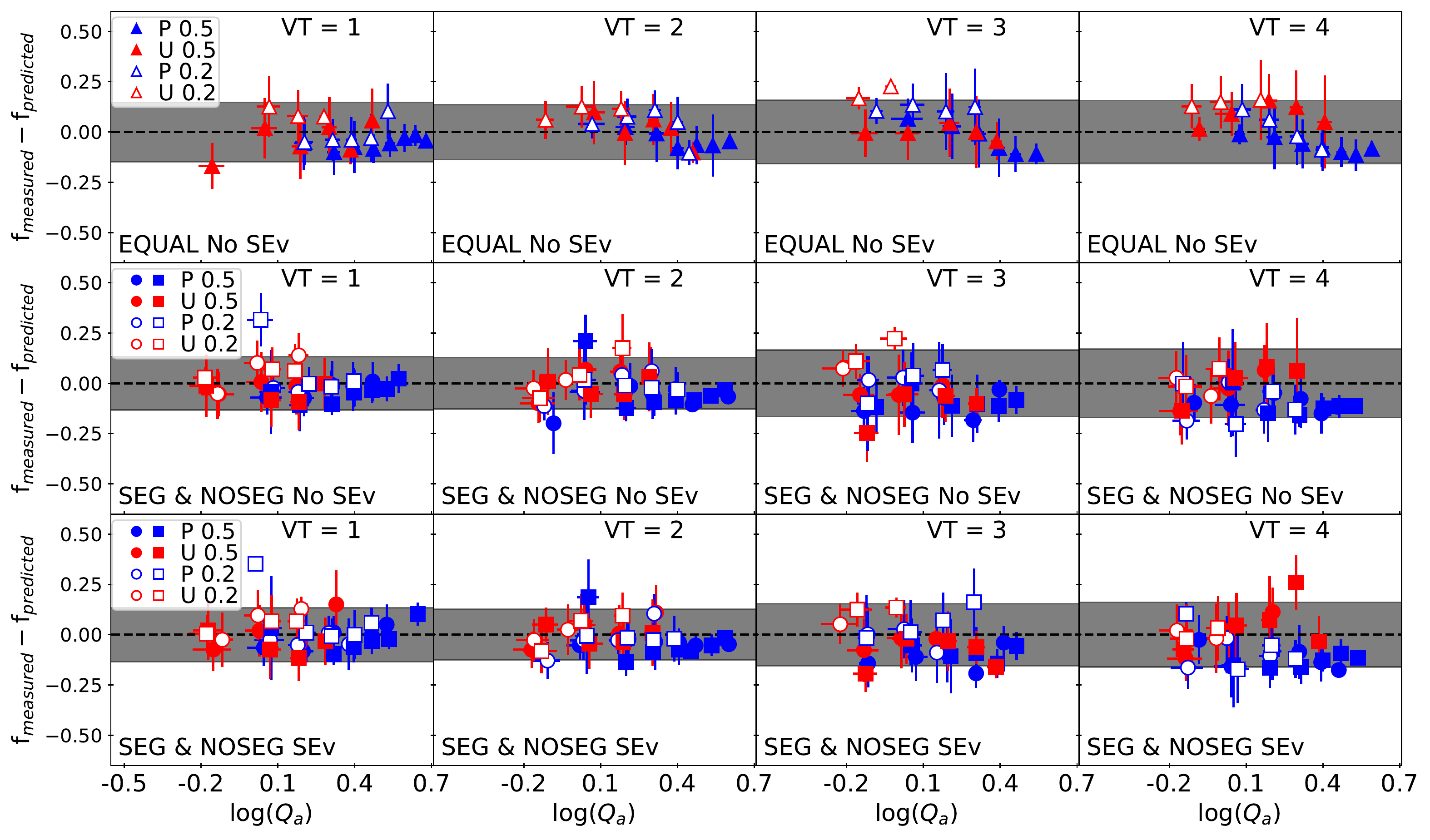}
    \caption{Same as Fig. \ref{fig:fbdifall16myrp2} but now for the moment of the gas expulsion.}
    \label{fig:fbdifallTEXPp2}
\end{figure*}

\begin{figure*}
    \centering
    \includegraphics[width=\textwidth]{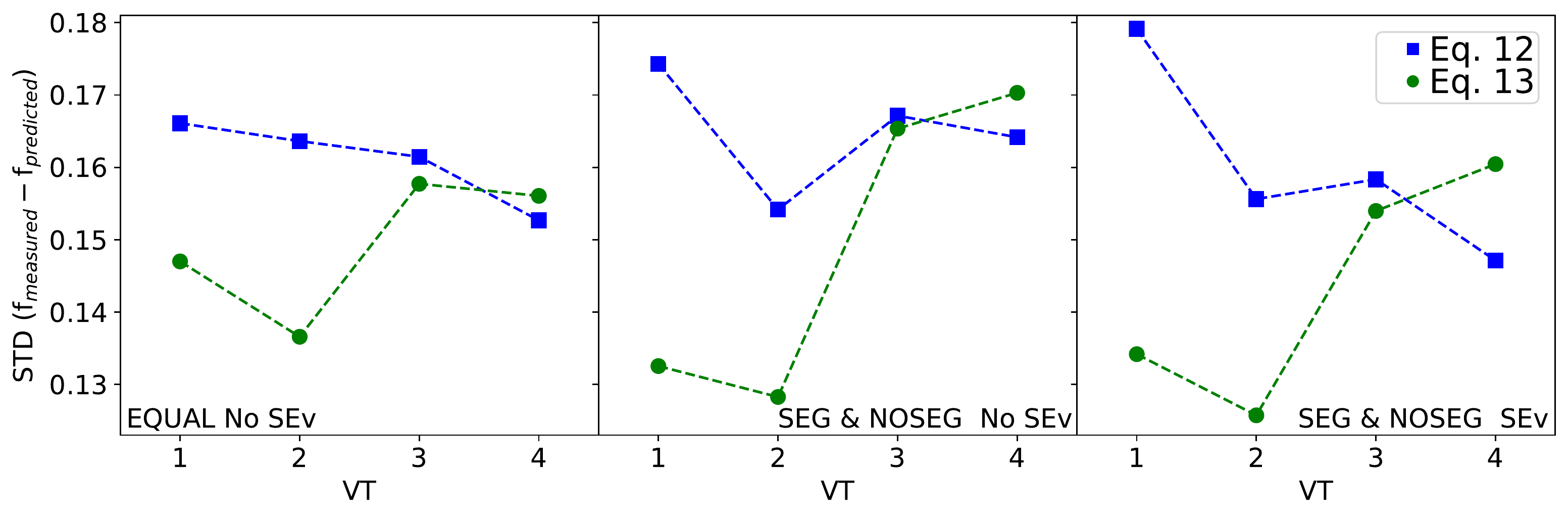}
    \caption{Same as in Fig. \ref{fig:avfbdiftexp16myr} but now \fbound is measured at the moment of the gas expulsion.} 
    \label{fig:avfbdiftexp16myr}
\end{figure*}

\subsection{Mass segregation}
\label{sec:ms}
We show in Fig. \ref{fig:tvsL} the time evolution of $\Lambda_\text{MSR}$. The symbols are the same as before and the solid black line is $\Lambda_\text{MSR}=1$ which means a stellar distribution where massive stars are distributed the same way as low mass stars, i.e., without mass segregation.

The left top panel is the evolution of mass segregation for simulations starting with SEG initial stellar distribution, and No-SEv. At the beginning, $\Lambda_\text{MSR}=4.05 \pm 0.77$ for all cases as they are the same fractals with the same IMF samples. At the moment of gas expulsion, which happens typically at a time of $ 1.18 \pm 0.50$ Myr, the level of mass segregation decreases until it reaches $\Lambda_\text{MSR}=3.02 \pm 1.49$. On the other hand, the simulations in the right panel at $t = 0$ Myr have $\Lambda_\text{MSR}=1.00 \pm 0.02$. At the moment of gas expulsion at $t = 1.25 \pm 0.55$ Myr the level of mass segregation increases to an average value of $\Lambda_\text{MSR}=1.32 \pm 0.35$. The same behaviour is observed in the bottom panels, where stellar evolution is activated. Initial values are the same as they are from the same clusters. Small differences can appear due to the randomness of choosing the sample of low mass stars for the calculation of $\Lambda_\text{MSR}$ (see \S~\ref{sec:lambdamsr}). The evolution of the virial ratio is very similar since only SEv mass loss from winds acts on timescales $\lesssim 1$ Myr for most of the stars. In the left panel at $t=1.19 \pm
0.50$ the mass segregation on average is $\Lambda_\text{MSR}=3.01 \pm 1.48$ and in right panel at 
$t=1.25$ $\pm$ $0.55$ it is $\Lambda_\text{MSR}=1.32 \pm 0.37$. The values for each of the cases are summarized in Tab.~\ref{tab:Lamdasum}. The gas expulsion for simulations with uniform BG potential is slightly later than simulations with Plummer BG potential.

As reported in \citet{Dom2017} and \citet{2019A&A...626A..79P}, we see that clusters with initial mass segregation rapidly decrease their levels during the embedded phase, as they relax into a more stable configuration.
On the other hand, clusters with NOSEG initial stellar distribution start 
raising their mass segregation levels until gas-expulsion happens.

\begin{table}
    \centering
    \caption{Summary of $\Lambda_\text{MSR}$. First and second columns indicate the BG potential (P = Plummer, U = uniform) with its respective initial virial ratio. The third column refers to the sample initial stellar distribution and if the stellar evolution is on. The fourth column shows the initial $\Lambda_\text{MSR}$ value and fifth column is the average time when gas expulsion is done. Sixth column is the $\Lambda_\text{MSR}$ at the moment of gas expulsion and last column shows the $\Lambda_\text{MSR}$ at 4.8 Myr.}
    \label{tab:Lamdasum}
    \resizebox{\columnwidth}{!}{
    \begin{tabular}{c|c|c|c|c|c|c}
         \hline
         BG & $Q_\text{i}$& Sample/SEv& Initial $\Lambda_\text{MSR}$ &Time TEXP &$\Lambda_{\text{MSR,ge}}$ &
         $\Lambda_{\text{MSR}, t = 4.8}$\\\hline
P & 0.5 & SEG/No & $4.05 \pm 0.77$ & $1.08 \pm 0.49$ & $3.01 \pm 1.36$ & $3.12 \pm 1.61$
\\
P & 0.2 & SEG/No & $4.05 \pm 0.76$ & $1.02 \pm 0.43$ & $3.62 \pm 1.46$ & $3.49 \pm 2.19$
\\
P & 0.5 & SEG/Yes & $4.06 \pm 0.77$ & $1.09 \pm 0.49$ & $2.96 \pm 1.35$ & $3.04 \pm 1.60$
\\
P & 0.2 & SEG/Yes & $4.05 \pm 0.76$ & $1.02 \pm 0.43$ & $3.56 \pm 1.49$ & $3.20 \pm 2.00$
\\ [4 pt]
U & 0.5 & SEG/No & $4.05 \pm 0.76$ & $1.22 \pm 0.49$ & $2.47 \pm 1.29$ & $3.01 \pm 1.78$
\\
U & 0.2 & SEG/No & $4.05 \pm 0.77$ & $1.41 \pm 0.49$ & $2.98 \pm 1.62$ & $2.55 \pm 2.05$
\\
U & 0.5 & SEG/Yes & $4.05 \pm 0.76$ & $1.23 \pm 0.49$ & $2.55 \pm 1.30$ & $3.14 \pm 1.88$
\\
U & 0.2 & SEG/Yes & $4.05 \pm 0.77$ & $1.41 \pm 0.49$ & $2.96 \pm 1.61$ & $2.44 \pm 1.98$
\\ [4 pt]
P & 0.5 & NOSEG/No & $1.00 \pm 0.02$ & $1.12 \pm 0.55$ & $1.16 \pm 0.25$ & $1.27 \pm 0.43$
\\
P & 0.2 & NOSEG/No & $1.00 \pm 0.03$ & $1.08 \pm 0.45$ & $1.30 \pm 0.29$ & $1.40 \pm 0.53$
\\
P & 0.5 & NOSEG/Yes & $1.00 \pm 0.02$ & $1.12 \pm 0.55$ & $1.16 \pm 0.26$ & $1.26 \pm 0.36$
\\
P & 0.2 & NOSEG/Yes & $1.00 \pm 0.03$ & $1.08 \pm 0.44$ & $1.31 \pm 0.31$ & $1.41 \pm 0.50$
\\ [4 pt]
U & 0.5 & NOSEG/No & $1.00 \pm 0.03$ & $1.28 \pm 0.58$ & $1.30 \pm 0.35$ & $1.45 \pm 0.42$
\\
U & 0.2 & NOSEG/No & $1.00 \pm 0.02$ & $1.51 \pm 0.51$ & $1.51 \pm 0.41$ & $1.77 \pm 0.83$
\\
U & 0.5 & NOSEG/Yes & $1.00 \pm 0.02$ & $1.28 \pm 0.59$ & $1.28 \pm 0.32$ & $1.43 \pm 0.39$
\\
U & 0.2 & NOSEG/Yes & $1.00 \pm 0.02$ & $1.51 \pm 0.52$ & $1.54 \pm 0.46$ & $1.86 \pm 0.95$
        \\\hline
    \end{tabular}}
\end{table}

We include a linear fit for times $\geq4.8$ Myr to each pair of initial conditions with the form:
\begin{eqnarray}
\Lambda_{\rm MSR}(t) &=& m_\Lambda t + \Lambda_{\text{MSR}, t = 4.8}, \ \ \ \ \ t \geq 4.8, 
\label{eq:fitL}
\end{eqnarray}
where  $m_\Lambda$ is the slope of the fit in units of Myr$^{-1}$ and $\Lambda_{\text{MSR}, t = 4.8}$ is the $\Lambda_\text{MSR}$ at 4.8 Myr, which is shown in Tab.~\ref{tab:Lamdasum}, last column.
As the clusters expand after gas expulsion, the value of $\Lambda_\text{MSR}$ shows a continuous decrease ($m_\Lambda<0$), being steeper for SEG simulations. Simulations with SEv are shown in the bottom panels. In this case, the decrease of $\Lambda_\text{MSR}$ is steeper as the more massive stars explode as SNe.

\subsection{Dynamical evaporation}
\label{sec:evap}
As we have introduced different masses, stronger interactions between the stars are expected, leading to the ejection of stars.
In addition, SEv adds another source of mass loss. In Fig.~\ref{fig:tvsFB} we show the \fbound evolution using the same symbols as in the previous figures. 
We include a linear fit to each pair of initial conditions with the form:
\begin{eqnarray}
f(t) &=& m_f t + f_{\rm bound,ge}, \ \ \ \ \ t \geq {\rm TEXP},
\label{eq:fitevap}
\end{eqnarray}
where  $m_f$ is the slope of the fit 
in units of Myr$^{-1}$ and $f_{\rm bound,ge}$ is the bound fraction at the moment of gas expulsion. A summary table of both parameters is shown in Table~\ref{tab:fbsum}.
While all measurements are taken at the same times, i.e., at $t=4.8$, 6.4, 8, 9.6, 11.2, 12.8, 13.4 and 16 Myr, they are slightly shifted for clarity.

In the top panel, the time evolution of \fbound is shown for simulations with equal-mass particles (triangles). For the four cases are observed practically constants \fbound 
values, as we measure two slopes with $m_f = -0.001$ for Plummer BG (P) and two slopes with $m_f = 0$ for uniform BG
(U).

In the panels where simulation with No-SEv are shown, we observe in both cases negative slopes with values $-0.003 \leq m_f \leq-0.009$ independent on the initial conditions. SEG simulations under uniform BG potential with $Q_\text{i}=0.2$ show the steepest slope.

In the bottom panels, where simulations with SEv are shown, we observe even steeper slopes with values $-0.005 \leq m_f \leq-0.011$ also independent of the initial conditions. As in the central panels, we measure the steepest slope in the left panel under the same initial conditions.


\begin{table}
    \centering
    \caption{Summary of fitting line slopes for \fbound time evolution. First and second columns indicate the BG potential (P = Plummer, U = uniform) with its respective initial virial ratio. The third column refers to the sample initial stellar distribution and if the stellar evolution is on. The fourth column shows the slope of the fitting line measured for each case. Last column shows \fbound at the moment of gas expulsion.}
    \label{tab:fbsum}
    \begin{tabular}{c|c|c|c|c}
         \hline
         BG & $Q_\text{i}$& Sample/SEv& $m_f$ & $f_{\rm bound,ge}$ 
         \\\hline
         P & 0.5&EQUAL/No &$-0.003$ & $ 0.23 \pm 0.16 $   \\
         P & 0.2&EQUAL/No&$-0.006$  & $ 0.44 \pm 0.18 $    \\[2pt]
         U & 0.5&EQUAL/No &$-0.003$ & $ 0.48 \pm 0.18 $   \\
         U & 0.2&EQUAL/No&$-0.006$  & $ 0.74 \pm 0.17 $    \\[6pt]
         P & 0.5&SEG/No &$-0.003$ & $ 0.29 \pm 0.14 $   \\
         P & 0.2&SEG/No&$-0.006$  & $ 0.43 \pm 0.16 $    \\[2pt]
         U & 0.5&SEG/No &$-0.005$ & $ 0.50 \pm 0.14 $     \\
         U & 0.2&SEG/No &$-0.009$ & $ 0.62 \pm 0.14 $     \\[4pt]
         P & 0.5&NOSEG/No&$-0.003$ & $ 0.22 \pm 0.14 $    \\
         P & 0.2&NOSEG/No &$-0.005$ & $ 0.35 \pm 0.17 $     \\[2pt]
         U & 0.5&NOSEG/No &$-0.003$  & $ 0.42 \pm 0.15 $    \\
         U & 0.2&NOSEG/No& $-0.007$  & $ 0.59 \pm 0.16 $    \\[6pt]
         P & 0.5&SEG/Yes &$-0.006$ & $ 0.25 \pm 0.13 $     \\
         P & 0.2&SEG/Yes &$-0.009$ & $ 0.38 \pm 0.17 $     \\[2pt]
         U & 0.5&SEG/Yes &$-0.008$ & $ 0.46 \pm 0.15 $     \\
         U & 0.2&SEG/Yes &$-0.011$ & $ 0.59 \pm 0.14 $    \\[4pt]
         P & 0.5&NOSEG/Yes &$-0.005$ & $ 0.18 \pm 0.14 $     \\
         P & 0.2&NOSEG/Yes &$-0.007$ & $ 0.32 \pm 0.16 $    \\[2pt]
         U & 0.5&NOSEG/Yes&$-0.006$  & $ 0.39 \pm 0.14 $    \\
         U & 0.2&NOSEG/Yes&$-0.010$  & $ 0.55 \pm 0.16 $
         \\\hline
    \end{tabular}
\end{table}

The highest values of \fbound are shown in all cases at the moment of gas expulsion, and continuously decrease thereafter (SEG-NOSEG). In Fig. \ref{fig:fbdifallTEXPp1} and Fig. \ref{fig:fbdifallTEXPp2}, we show again the average difference with the prediction but now compared with the value of \fbound measured at the moment of gas expulsion. At this moment, both Eq.~\ref{eq:model1} and Eq. \ref{eq:model2} can closely describe our results, with the first prediction, compared with equal-mass results, still showing larger dispersion for early VT, but in the same range for later gas expulsion and, this is independent of the inclusion or not of SEv. We summarize in Tab. \ref{tab:fbsum} the different fitting line slopes together with their respective \fbound at the moment of gas expulsion. We find that independent of the intrinsic characteristics in our sample, 
they show decreasing slopes, as stars are ejected.
On a first order, models with a Plummer background potential have stronger slopes than uniform background potential.
And to second order, star clusters with $Q_i=0.5$ have steeper slopes than initially cold star clusters.

\section{Summary \& Conclusion}
\label{sec:sumdisc}
In this work, we test two models introduced by \citet{Farias2015} and \citet{Farias2018}, that predict the fraction of bound mass that star clusters can retain after explosive gas expulsion. These models were previously tested only using equal mass particle and fractal star clusters. Here, we explore how these models work on a more realistic scenario, introducing an IMF with two different particle distribution for the primordial location of massive stars. We first assume massive stars are born in random locations within the star cluster (NOSEG), and models with high levels of primordial mass segregation (SEG). In both cases, we also investigate the effects of SEv. We create a sample of 800 simulations to minimize stochastic fluctuations for every set and combination of the new parameters.

Since we use the latest version of the Nbody6++ code, we start by reproducing the previous results with equal-mass particles. In order to be consistent with recent evidence of a very early release of gas in low mass clusters $\sim$ 1 Myr \citep{2020MNRAS.499..748D}, we set the moment of gas expulsion at an early time. 

The first predictive model, which depends on the local stellar fraction (LSF) and the pre-gas expulsion virial ratio, we find that the results are more accurate when star clusters expel their gas at later stages, i.e., when the level of substructure is reduced by dynamical processes, in agreement with our previous works. The second predictive model, that only depends of one parameter, the post-gas expulsion virial ratio, is not sensible to substructure, confirming previous results tested with highly substructured background gas models, and indicating that the nature of the background gas makes no difference for this specific model.

We introduce random IMF samples with different levels of primordial mass segregation as quantified by $\Lambda_\text{MSR}$ and contrast with previous work.
Star clusters with no primordial mass segregation, show a lower concentration of massive stars at the moment of gas expulsion compared to clusters with primordial segregation. 
Non segregated star clusters are still raising their concentration of stars when gas is removed, as reported in previous work using similar frameworks \citep[see e.g..][]{2010MNRAS.407.1098A,2011ApJ...732...16Y,Dom2017}. We find an average $\Lambda_\text{MSR} \sim 1.32$ implying that all clusters in this work are mass segregated at the moment of gas expulsion, regardless of the initial conditions. 

After gas expulsion, $\Lambda_\text{MSR}$ is observed for a short time to be even higher, followed by a continuous decrease due to cluster expansion. For the case of simulations with SEv, the SNe occurring in the second half of the simulation results in a steeper decrease. 

By introducing random IMF samples, \fbound measurements at 16 Myr are in most of the cases below the predicted curves and with larger deviations when SEv is included. By examining the evolution in time of \fbound, we observe a continuous decrease or a negative slope, also known as dynamical evaporation. The average values of \fbound are much closer to the predictions when they are measured at the moment of gas expulsion. For equal-mass particles the evaporation slopes are close to be zero, i.e., the predictions from the model are matched independent of the time we measure \fbound as it stays practically constant.

For simulations with SEv, at the moment of gas expulsion, we observe a similar trend. While SEv mass loss should decrease the bound mass, our simulations show that this is not the case at early times. Our low mass clusters contain only a few very massive stars ($M>20$ M$_\odot$) and only stellar winds change the mass at an early phase. These few massive stars evolve as SNe in the second part of the simulation and thus we only observe differences at later times. 

We conclude that independent of the initial conditions, the predictive analytical models introduced by \citet{Farias2015,Farias2018} can describe our results when measuring close to the gas expulsion time, but they overestimate \fbound at later stages. Dynamical evaporation due to two-body interactions is stronger in stellar systems having different stellar masses and it is the main reason for the continuous decrease of \fbound. The inclusion of SEv can only decrease \fbound at later stages due to SNe mass loss. Moreover, no significant differences are observed at early times when only stellar winds take action. Initial mass distribution (SEG or NOSEG) does not play a role in our results. This is due to the fact that all clusters studied have rapidly become dynamically mass segregated, irrespective of the details of the initial conditions. We emphasize that the gas expulsion scheme studied here is the most destructive scenario and any smoothing applied to the process would improve the chances to find larger \fbound. 

Whether or not our initial conditions are a realistic state of an embedded star cluster is a matter of discussion. It has been shown that the pre-gas expulsion of a young massive star cluster ($M > 10^4$ M$_\odot$) is very compact \citep{2012A&A...543A...8M} and the number of substructures weakly depend on their total mass. An example of this is R136 in the 30 Doradus nebula with a total mass $> 2.2 \times 10^4$ M$_\odot$ with a radius poorly constrains to be $\sim 2$ pc. Authors \citep[see e.g.][]{2017MNRAS.465.1375S} also suggest that this object is not result of a single starburst and probably a (re)-collapse of gas which already gave birth to an older generation of stars where both together are part of NGC2070, and this scenario differs with the one developed here. In this work, the low mass embedded star clusters ($M = 2.5 \times 10^3$ M$_\odot$) show substructures at the moment of gas expulsion and these are decreasing as we wait to remove the gas. The aim of this work is to study if this non-spherical distribution helps the cluster to deal better with the violent gravitational potential change and to survive with SFE = 0.20 otherwise if we wait until the substructures are erased we would reproduce the same scenario largely study in \citet{2007MNRAS.380.1589B} with non-surviving clusters in this range of SFE.

\section*{Acknowledgments:} We acknowledge support from CONICYT (CONICYT-PFCHA/Doctorado acuerdo bilateral DAAD/62170008), financial support from DAAD (funding program number 57395809). The authors acknowledge support by the state of Baden-W\"urttemberg through bwHPC and the German Research Foundation (DFG) through grant INST 35/1134-1 FUGG and Heidelberg cluster of excellence EXC 2181 (Project-ID 390900948) "STRUCTURES: A unifying approach to emergent phenomena in the physical world, mathematics, and complex data" funded by the German Excellence Strategy. JPF acknowledges support from ERC Advanced Grant MSTAR. MF acknowledges funding from FONDECYT regular N$^\text{o}$ 1180291, BASAL N$^\text{o}$ AFB-170002 (CATA). R.S.K. acknowledges financial support from the DFG via the collaborative research center SFB 881 ``The Milky Way System'' (project ID 138713538, subprojects B1, B2, B8).

\section*{DATA AVAILABILITY STATEMENT}
The data of the full set of simulations (see Tab. \ref{tab:initcond}) presented in this article will be shared on reasonable request to the corresponding author.

\bibliographystyle{mnras}
\bibliography{bibtex} 

\label{lastpage}

\end{document}